\documentclass[aps,prc,twocolumn,superscriptaddress]{revtex4-1}
\usepackage{graphicx}
\usepackage{amsmath}
\usepackage{subfigure}
\usepackage{subfigure,dcolumn}
\usepackage{appendix}
\usepackage{xcolor}
\usepackage{ulem}
\usepackage{enumerate}
\usepackage{enumitem}
\usepackage[colorlinks,
linkcolor=blue,
anchorcolor=blue,
urlcolor=red,
citecolor=blue]{hyperref}
\begin{document}
	

	\title{Temperature dependence of the nucleon–nucleon inelastic cross section in an isospin-asymmetric nuclear medium}

	

\author{Manzi Nan}
\affiliation{School of Science, Huzhou University, Huzhou 313000, China}
\affiliation{Institute of Modern Physics, Chinese Academy of Sciences, Lanzhou 730000, China}
\affiliation{School of Nuclear Science and Technology, University of Chinese Academy of Sciences, Beijing 100049, China}
\affiliation{School of Nuclear Science and Engineering, Shenyang Institute of Engineering, Shenyang 110136, China}
\author{Pengcheng Li}
\email[Corresponding author, ]{lipch@zjhu.edu.cn}
\affiliation{School of Science, Huzhou University, Huzhou 313000, China}
\author{Guojun Wei}
\affiliation{School of Science, Huzhou University, Huzhou 313000, China}
\author{Xilong Xiang}
\affiliation{School of Science, Huzhou University, Huzhou 313000, China}
\author{Wei Zuo}
\affiliation{Institute of Modern Physics, Chinese Academy of Sciences, Lanzhou 730000, China}
\affiliation{School of Nuclear Science and Technology, University of Chinese Academy of Sciences, Beijing 100049, China}
\author{Qingfeng Li}
\affiliation{School of Science, Huzhou University, Huzhou 313000, China}
\affiliation{Institute of Modern Physics, Chinese Academy of Sciences, Lanzhou 730000, China}
 
\date{\today}

\begin{abstract}		
The nucleon–nucleon ($NN$) inelastic cross section plays an important role in constraining the nuclear equation of state at high baryon density and in describing the formation and evolution of compact astrophysical objects. 
In this study, the temperature $T$ dependence of the $\Delta^{++}$ and $\Delta^{-}$ production cross sections in the isospin-symmetric and -asymmetric nuclear medium is investigated within the self-consistent and relativistic Boltzmann–Uehling–Uhlenbeck (RBUU) framework. 
Two relativistic mean-field parameterizations are employed: the density-dependent parameterization (called DD-ME$\delta$) and the nonlinear-dependent parameterization (called OMEG). 
Both parameterizations yield similar $T$-dependent baryon effective masses and mass splittings, although the OMEG set exhibits a stronger density dependence, particularly at higher densities ($>  1.5\rho_{0}$).
Consequently, at lower densities, the energy, density, temperature, and isospin dependence of both $\Delta^{++}$ and $\Delta^{-}$ production cross sections are comparable for both sets, whereas at higher densities, the OMEG set predicts a stronger temperature and density sensitivity. 
Moreover, the $T$ dependence of the $NN$ inelastic cross section is enhanced with increasing density, but is suppressed in isospin-asymmetric nuclear matter compared to that in isospin-symmetric nuclear matter. 
The isospin dependence of the cross section remains nearly $T$-independent at small asymmetries, yet becomes more intricate in highly asymmetric systems. 
These findings provide valuable testing inputs for improving the thermal treatment of $\Delta$ related dynamical processes in transport models and offer insights into the behavior of $\Delta$ in astrophysical environments, such as core-collapse supernovae and binary neutron star mergers.
\end{abstract}


\maketitle
	
\section{Introduction}\label{sec:1}

The $\Delta$(1232) plays a special role in understanding the dynamics of intermediate-energy heavy-ion collisions (HICs) and various astrophysical phenomena \cite{Teis:1996kx,Li:1995pra,Stoecker:1986ci,Cai:2015hya,Drago:2014oja}. 
In intermediate-energy HICs, the $\Delta$ resonance is the most abundant baryonic resonance due to its relatively low excitation energy. 
And the key reaction channels involved include but are not limited to nucleon-nucleon inelastic scattering ($NN \rightarrow N\Delta$), pion absorption ($N\pi \rightarrow \Delta$), and $\Delta$ decay ($\Delta \rightarrow N \pi$), as well as the nucleon-$\Delta$ elastic scattering ($N\Delta \rightarrow N\Delta$) \cite{Huber:1994ee,TerHaar:1986xpv,Mao:1994zza}. 
Further, the $\pi$-related observables in HICs and the extracted information of the high-density nuclear equation of state (EoS) using these observables will be directly influenced by the production and evolution of the $\Delta$ particles \cite{Li:2005gfa,Xiao:2008vm,TMEP:2019yci,TMEP:2023ifw}.  
Additionally, in astrophysics, recent studies have revealed that $\Delta$ particles emerge at densities around 2$\rho_{0}$ in neutron stars \cite{Cai:2015hya}, contrasting with earlier predictions of significantly higher onset densities \cite{Glendenning:1984jr,Glendenning:1998zx}. 
The appearance of $\Delta$ resonances ought to significantly soften the EoS, thereby altering the mass-radius relation and internal composition of neutron stars \cite{Cai:2015hya,Drago:2014oja}, as well as the supernova dynamics and the binary neutron star mergers \cite{Sedrakian:2022kgj}.

Over the past three decades, as the daughter nuclei of $\Delta$, the $\pi$ and its related observables, such as the $\pi^-/\pi^+$ ratio, have often been used to constrain the density dependence of the symmetry energy \cite{Li:2002qx,Chen:2004si,FOPI:2006ifg,SpiRIT:2020sfn}. 
However, based on the comparison of the simulated results of the transport models with experimental data, the derived density dependence of the symmetry energy, especially at high densities, still exhibits large uncertainties, even the contradictory trend \cite{Guo:2014tua,Li:2019xxz}. 
In addition, while different transport models yield inconsistent predictions for the rapidity and transverse momentum distributions and fail to reproduce the experimental trends, they also systematically overpredict the pion multiplicities, a long-standing discrepancy that has persisted for more than two decades \cite{TMEP:2022xjg,Bass:1995pj,Larionov:2001va, Larionov:2003av,UmaMaheswari:1997ig}. 
A growing body of evidence suggests that the discrepancy originates also partly from insufficient treatment of in-medium modifications at least to the $NN\rightarrow N\Delta$ cross section \cite{Liu:2001iz,Godbey:2021tbt,Kim:2022sbj,Kummer:2023hvl,Li:2025uku,Guo:2025xie,Han:2025hqm}. 
For example, in Ref.~\cite{Godbey:2021tbt}, by introducing a density-dependent reduction factor within the relativistic Vlasov-Uehling-Uhlenbeck (RVUU) model, they obtained a better description of the $\pi$ rapidity distributions and transverse momentum spectra of the HADES Collaboration \cite{HADES:2020ver}. 
However, this approach remains essentially phenomenological, and a fully self-consistent framework is urgently required, in which the mean-field potential and the collision term are derived from the same underlying interaction, properly accounting for their energy, density, isospin, and temperature dependence. 

In order to achieve a more self-consistent treatment of the mean-field potential and the collision term within a unified framework, several approaches have been developed. One class is based on Dirac–Brueckner–Hartree–Fock (DBHF) calculations, in which the scalar and vector self-energies are obtained from the DBHF $G$-matrix \cite{Gaitanos:1998bu}, and implemented as the input for transport equations. A second direction is the $\chi$BUU framework, which incorporates interactions derived from chiral effective field theory, however, in this formulation only the mean-field potentials are treated covariantly \cite{Zhang:2018ool}. A third approach adopts a well-established phenomenological strategy based on the Quantum Hadrodynamics (QHD), where the scalar and vector fields arise from effective Lagrangians calibrated to empirical nuclear matter properties \cite{Walecka:1974qa}.
In our previous work, based on the relativistic Boltzmann–Uehling–Uhlenbeck (RBUU) theory framework with an effective Lagrangian including the $\sigma$, $\omega$, $\delta$, $\rho$, and $\pi$ mesons coupling to both nucleons and $\Delta$ resonances, the energy, density, and isospin dependence of the 
$NN \rightarrow NN$ \cite{Mao:1994et, Li:2000sha,Li:2003vd}, 
$NN \rightarrow N\Delta$ \cite{Li:2016xix},
$N\pi \rightarrow \Delta$ \cite{Li:2017pis},  
and $N\Delta \rightarrow N\Delta$ \cite{Nan:2023gwp,Nan:2024ogc} cross sections can be investigated meanwhile. 
It was found that these reactions exhibit rich degrees of energy, density, and isospin dependence, where the isospin effect is dominantly contributed by the isovector $\delta$ and $\rho$ meson exchanges. 
These theoretical endeavors are definitely helpful to make more reliable conclusions on the high-density symmetry energy. 

In addition, the particle energy spectra are usually used to extract the effective temperature in HICs \cite{Gaitanos:2003zg,Reichert:2022qys,Reichert:2020uxs,Brockmann:1984de,Wang:1996zz,Andronic:2005yp},  
but, the influence of the temperature effect on the cross sections is frequently neglected, which should be received more attention. 
And, as the density increases, the temperature effect will become obvious, especially for the isospin asymmetric systems \cite{Li:2003vd,Sen:2020edi}. 
In this work, by adopting the same self-consistent RBUU theory framework as that established in our previous work \cite{Li:2000sha,Li:2003vd,Li:2016xix,Li:2017pis,Nan:2023gwp,Nan:2024ogc}, the microscopic calculation of the $NN \rightarrow N\Delta$ cross section is further extended to include temperature dependence. 

The paper is organized as follows. Section~\ref{sec2} briefly introduces the effective Lagrangian and the RBUU approach with finite temperature. In Section~\ref{sec3}, the numerical results for the temperature-dependent $NN \rightarrow N\Delta$ cross sections are shown and discussed. Finally, the summary and outlook are provided in Section~\ref{sec4}.

\section{Formulation}\label{sec2}

We begin with the QHD-type Lagrangian. In the determination of exchange mesons within the Lagrangian, the isospin-vector $\delta$ meson exchange is further introduced in addition to the usual isospin-scalar $\sigma$, $\omega$, and isospin-vector $\rho$ meson exchanges, as it significantly affects the symmetry energy at high densities, the splitting of neutron and proton effective masses, as well as the dynamics of HICs \cite{Gaitanos:2003zg}. 
The effective Lagrangian of the system can be divided into the free part and the interaction part, 
\begin{equation}
		\mathcal{L}=\mathcal{L}_{F}+\mathcal{L}_{I}.
\end{equation}

The free Lagrangian density $\mathcal{L}_{F}$ can be expressed as
\begin{equation}
    \begin{aligned}
\mathcal{L}_{F}= & \bar{\Psi}\left[i \gamma_{\mu} \partial^{\mu}-m_{N}\right] \Psi+\bar{\Psi}_{\Delta \nu}\left[i \gamma_{\mu} \partial^{\mu}-m_{\Delta}\right] \Psi_{\Delta}^{\nu} \\
& +\frac{1}{2} \partial_{\mu} \sigma \partial^{\mu} \sigma+\frac{1}{2} \partial_{\mu} \vec{\delta} \partial^{\mu} \vec{\delta}-\frac{1}{4} F_{\mu \nu} \cdot F^{\mu v}-\frac{1}{4} \vec{L}_{\mu\nu} \cdot \vec{L}^{\mu \nu} \\
& -\frac{1}{2} m_{\sigma}^{2} \sigma^{2}-\frac{1}{2} m_{\delta}^{2} \vec{\delta}^{2} +\frac{1}{2} m_{\omega}^{2} \omega_{\mu} \omega^{\mu}+\frac{1}{2} m_{\rho}^{2} \vec{\rho}_{\mu} \vec{\rho}^{\mu}\\
& -U_{NL}(\sigma,\omega^{\mu},\vec{\delta}, \vec{\rho}^{\mu}),
\end{aligned}
\end{equation}
where $F_{\mu \nu} \equiv \partial_{\mu} \omega_{v}-\partial_{v} \omega_{\mu}, 
\vec{L}_{\mu\nu} \equiv \partial_{\mu} \vec{\rho}_{v}-\partial_{v} \vec{\rho}_{\mu} $, 
$\Psi$ is the Dirac spinor of the nucleon, 
$\Psi_{\Delta}$ is the Rarita-Schwinger spinor of the $\Delta$ resonance. 
And $U_{NL}(\sigma,\omega^{\mu},\vec{\delta}, \vec{\rho}^{\mu})$ represents the nonlinear potential associated with meson–meson couplings,
\begin{equation}
    \begin{aligned}
        U_{NL}(\sigma,\omega^{\mu},\vec{\delta}, \vec{\rho}^{\mu})=& \frac{1}{3}g_{2}\sigma^{3}+\frac{1}{4}g_{3}\sigma^{4}\\
        -&\Lambda_{s}({g_{NN}^{\sigma}}^2 \sigma^{2})({g_{NN}^{\delta}}^2 \vec{\delta}^{2})\\
        -&\Lambda_{\nu}({g_{NN}^{\omega}}^2 \omega_{\mu}\omega^{\mu})({g_{NN}^{\rho}}^2 \vec{\rho}_{\mu}\cdot \vec{\rho}^{\mu}).
    \end{aligned}
\end{equation}
In the case of density-dependent nucleon–meson couplings, the nonlinear potential $U_{NL}(\sigma,\omega^{\mu},\vec{\delta}, \vec{\rho}^{\mu})$ is omitted.

The interaction part of baryons coupled to mesons $\mathcal{L}_{I}$, can be expressed as
\begin{equation}
    \begin{aligned}
\mathcal{L}_{I}= & g_{N N}^{\sigma} \bar{\Psi} \Psi \sigma+g_{N N}^{\delta} \bar{\Psi}\vec{\tau }\cdot \Psi \vec{\delta} -g_{N N}^{\omega} \bar{\Psi} \gamma_{\mu} \Psi \omega^{\mu}\\
& -g_{N N}^{\rho} \bar{\Psi} \gamma_{\mu} \vec{\tau } \cdot \Psi \vec{\rho }^{\mu} +g_{\Delta \Delta}^{\sigma} \bar{\Psi}_{\Delta} \Psi_{\Delta} \sigma+g_{\Delta \Delta}^{\delta} \bar{\Psi}_{\Delta} \vec{\tau } \cdot \Psi_{\Delta} \vec{\delta} \\
&-g_{\Delta \Delta}^{\omega} \bar{\Psi}_{\Delta} \gamma_{\mu} \Psi_{\Delta} \omega^{\mu}-g_{\Delta \Delta}^{\rho} \bar{\Psi}_{\Delta} \gamma_{\mu} \vec{\tau} \cdot \Psi_{\Delta} \vec{\rho }^{\mu},
\end{aligned}
\end{equation}
where $\vec{\tau}$ is the nucleon isospin operator.

\begin{figure}[t]
    \centering
    \includegraphics[width=1.0\linewidth]{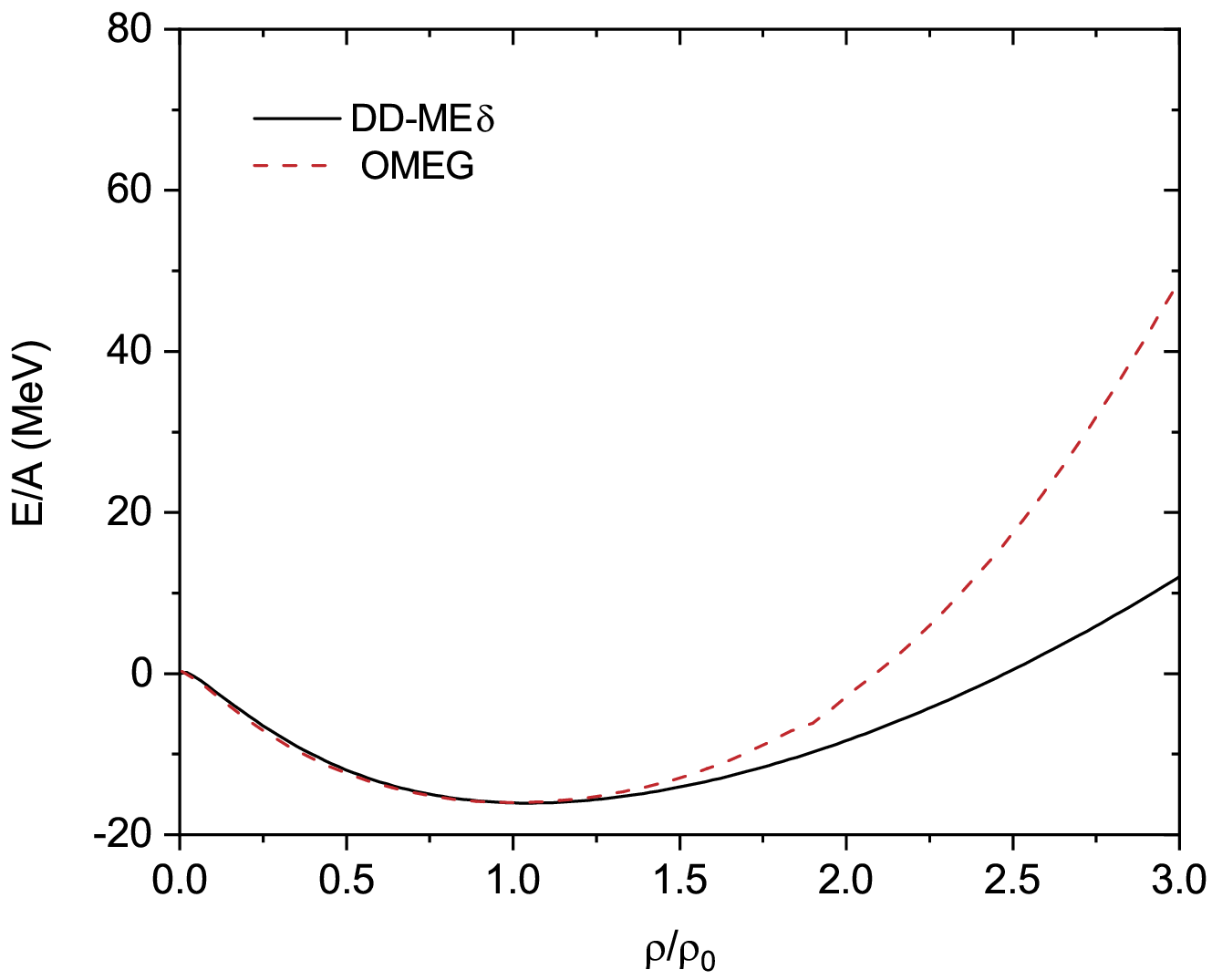}
   \caption{
   (Color online)
   Density-dependent EoS for symmetric nuclear matter. The black solid and red dotted lines denote the results for DD-ME$\delta$ and OMEG parametrizations, respectively. }
    \label{fig.0}
\end{figure}

Regarding the choice of parameter sets for the nucleon–nucleon–meson coupling constants, two distinct parameterizations are adopted and compared in order to show model dependence: the nonlinear-dependent coupling constants, called OMEG \cite{Miyatsu:2022wuy,Sun:2022yor}, which are frequently employed for astrophysical observations, and the density-dependent coupling constants, called DD-ME$\delta$ \cite{Roca-Maza:2011alv}, which are commonly used for terrestrial experiments with atomic nuclei.
In detail, for the OMEG parameter set, $g_{\sigma}$ = 9.22, $g_{\omega}$ = 11.35, $g_{2}$ = 13.08 $\text{fm}^{-1}$, and $g_{3}$ = -31.60 $\text{fm}^{-1}$ are adopted. 
The $g_{\delta}^2/4\pi$ = 2.488 is taken from the one-boson-exchange potential \cite{Machleidt:1989tm}, and $\Lambda_{\sigma\delta}$ = 50 \cite{Zabari:2018tjk,Zabari:2019clk}, 
while the $g_{\rho}^2/4\pi$ = 3.39 and $\Lambda_{\omega\rho}$ = 173.77 are determined by the given $E_{sym}$ = 32.0 MeV and $L$ = 50 MeV constrained from the HIC experiments \cite{Miyatsu:2022wuy}. 
These two parameterizations lead to slightly different EoSs for symmetric nuclear matter, as shown in Fig.~\ref{fig.0}.
The corresponding incompressibilities at saturation density are $K_{0}=219$ MeV for DD-ME$\delta$ and $K_{0}=230$ MeV for OMEG, respectively.
More pronounced differences between the two EoSs emerge at supra-saturation densities, reflecting differences in their stiffness.
Such differences are expected to influence the in-medium baryon effective masses and consequently affect the in-medium modification of the $NN$ inelastic cross sections discussed below.

It should be emphasized that the relativistic mean-field theory employing the OMEG parameter set has been shown to reproduce the properties of astrophysical observations, such as the tidal deformability of binary neutron star mergers, as well as the properties of the compact binary coalescence event GW190814 \cite{Miyatsu:2022wuy}. 
Additionally, compared to the earlier DD-ME2 set, which does not include the $\delta$ meson exchange \cite{Roca-Maza:2011alv}, the DD-ME$\delta$ parameter set improves the accuracy of the properties of finite nuclei, such as masses and radii, and provides a more reliable description for high-density EoS \cite{Roca-Maza:2011alv}. 

In uniform matter, the equation of motion of the meson fields can be given by 
\begin{align}
\left(m_{\sigma}^{2} + g_{2}{\sigma} + g_{3}{\sigma}^{2} - 2\Lambda_{\sigma\delta}{\delta}^{2}\right){\sigma} 
&= g_{\sigma}\left(\rho_{p}^{s} + \rho_{n}^{s}\right), 
\label{eq.5}
\\
\left(m_{\omega}^{2} + 2\Lambda_{\omega\rho}{\rho}^{2}\right){\omega} 
&= g_{\omega}\left(\rho_{p} + \rho_{n}\right), \\
\left(m_{\delta}^{2} - 2\Lambda_{\sigma\delta}{\sigma}^{2}\right){\delta} 
&= g_{\delta}\left(\rho_{p}^{s} - \rho_{n}^{s}\right), \\
\left(m_{\rho}^{2} + 2\Lambda_{\omega\rho}{\omega}^{2}\right){\rho} 
&= g_{\rho}\left(\rho_{p} - \rho_{n}\right).
\end{align}
The baryon density $\rho_{i}$ and scalar densities $\rho_{i}^{s}$ are 
\begin{equation}
\rho_{i} =  2 \int \frac{d^3 \mathbf{p}}{(2\pi)^3} \left[ f_i(\mathbf{p}) - \bar{f}_i(\mathbf{p}) \right],
\end{equation}
\begin{equation}
    \rho_{i}^{s} =  2 \int \frac{d^3 \mathbf{p}}{(2\pi)^3} \frac{m_i^*}{E_i^*} \left[ f_i(\mathbf{p}) + \bar{f}_i(\mathbf{p}) \right],
\end{equation}
the index $i$ denotes the particle species, proton or neutron. The total baryon density $\rho=\rho_p+\rho_n$, The $f_i(\mathbf{p})$ and $\bar{f}_i(\mathbf{p})$ are fermion and anti-fermion distribution functions, respectively, expressed as
\begin{equation}
\begin{array}{l}
f_i(\mathbf{p}) = 1/(1 + e^{[E_i^*(\mathbf{p}) - \mu_i^*]/T}), \\
\bar{f}_i(\mathbf{p}) = 1/(1 + e^{[E_i^*(\mathbf{p}) + \mu_i^*]/T}),
\label{eq.11}
\end{array}
\end{equation}
where $T$ is the temperature, $\mu^*_i$ is the effective chemical potential, and $E_{i}^{*}=\sqrt{{p^2}+{m^*_{i}}^2}$.
In the Hartree approximation of the relativistic mean-field theory, the effective masses for protons $m^*_{p}$ and neutrons $m^*_{n}$ are determined by the average value of the $\sigma$ and $\delta$ exchanges, which reads as 
\begin{equation}
   m_{p / n}^{*}=m_{N}-g^{\sigma}_{NN} \sigma \mp g^{\delta}_{NN} \delta. \\
   \label{nucleon mass}
\end{equation} 
Combined with Eqs.~\ref{eq.5}–\ref{nucleon mass}, the  $\sigma$, $\omega$, $\rho$, and $\delta$ meson fields can be obtained through a self-consistent iterative procedure. 
It should be noted that the effective mass adopted in this work corresponds to the relativistic effective mass, i.e., the so-called Dirac mass,  which is defined through the scalar self-energy. This quantity should be distinguished from the effective $k$-mass, which characterizes the momentum dependence of the mean-field potential \cite{Yang:2025aia}.
In addition to the standard Hartree treatment, nonlinear meson couplings may further modify the high-density behavior. In particular, the $\sigma$–$\delta$ coupling $\Lambda_{\sigma\delta}$ was found in Ref.~\cite{Miyatsu:2022wuy} to soften the symmetry energy at supra-saturation densities, which may in turn affect the nucleon effective masses in the high-density region. The corresponding results are presented below.

\begin{figure}[!t]
    \centering
    \includegraphics[width=1.0\linewidth]{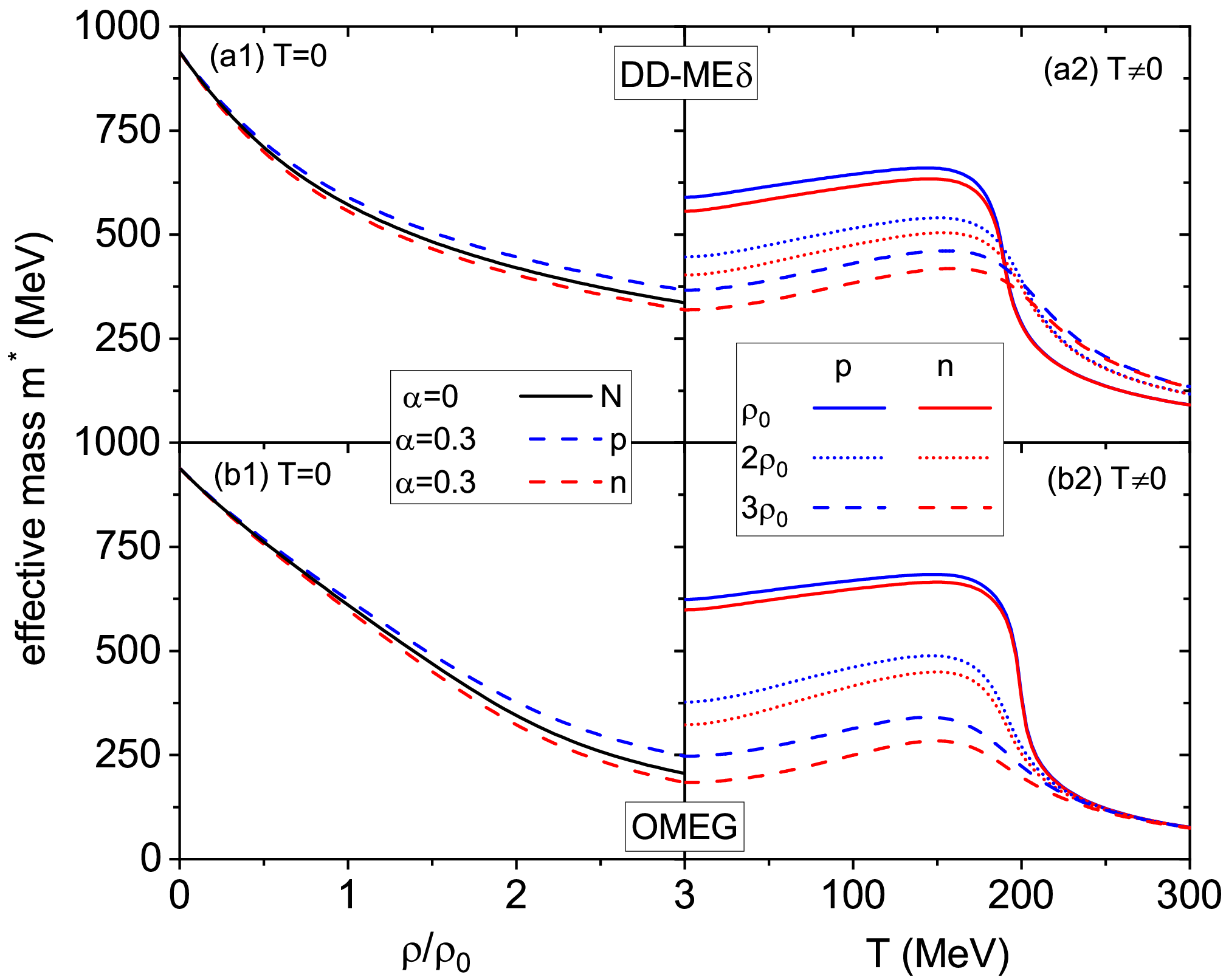}
   \caption{(Color online) Nucleon effective mass with respect to DD-ME$\delta$ set (top panels) and OMEG parameter set (bottom panels). The left panels depict the density dependence of nucleon effective mass at $\alpha$=0 (solid lines) and $\alpha$=0.3 (dashed lines) in cold nuclear matter. 
   The right panels show the temperature dependence of nucleon effective mass at densities of $\rho_{0}$ (solid lines), 2$\rho_{0}$ (dotted lines), 3$\rho_{0}$ (dashed lines) at $\alpha$=0.3. }
    \label{fig.1}
\end{figure}

The density and temperature dependence of the proton and neutron effective masses are presented in Fig.~\ref{fig.1}. 
The top panels display the results of calculations with the DD-ME$\delta$ parameter set, whereas the bottom panels show those obtained with OMEG set.
The left panels illustrate the density dependence of the baryon effective masses at $T=0$ for isospin asymmetries $\alpha=(\rho_n-\rho_p)/(\rho_n+\rho_p)=0$ and 0.3.
It can be seen that the proton effective mass remains consistently larger than that of the neutron, and this splitting, caused by $\delta$ meson exchange, becomes more pronounced with increasing density. 
It should be pointed out that one can obtain different orderings of the nucleon effective masses from different analyses, optical-potential studies generally favor $m_n^{*}>m_p^{*}$ \cite{Li:2004zi}, whereas HICs data support the opposite trend of $m_n^{*}<m_p^{*}$ \cite{Zhang:2014sva, Coupland:2014gya,Su:2017vwn}. 
And this discrepancy can be partly resolved by considering the momentum-dependent symmetry potential \cite{Yang:2025aia}. 
In addition, at lower densities ($\rho \le 1.5 \rho_0$), the DD-ME$\delta$ and OMEG results show similarity in describing the density dependence of the nucleon effective mass. However, at higher densities ($1.5 \rho_0 < \rho \le 3 \rho_0$), the results obtained using the OMEG set show a significantly steeper decrease compared to those obtained using the DD-ME$\delta$ set.

The right figures depict the temperature dependence of effective masses at $\rho$=$\rho_{0}$, 2$\rho_{0}$, 3$\rho_{0}$ for $\alpha = 0.3$.  
For the temperature range of 0$\sim$150 MeV, the effective masses slowly increased with temperature, while the effective mass splitting between protons and neutrons is weakly independent of temperature but strongly dependent on the density.  
These indicate that, in a hot nuclear medium, the temperature effect is predominantly caused by the $\sigma$ meson exchange, while the isospin effect is mainly contributed by the $\delta$ meson exchange, and the related cross sections would be more significantly affected in higher density regions, which will be further discussed in the following sections. 
For the temperature above 150 MeV, the effective masses sharply decrease with temperature, which is similar to the temperature dependence of the constituent quark mass caused by the chiral symmetry being restored in the Nambu–Jona-Lasinio (NJL) model \cite{Zhuang:1994dw,Chaudhuri:2019lbw}. 
And the decrease here is largely driven by particle–antiparticle pair production \cite{Theis:1983egm,Freedman:1977fd,Glendenning:1986ib}. 
This temperature region may also be associated with a possible phase transition from a hadronic gas to a quark–gluon plasma (QGP)\cite{Bellwied:2015rza}. Therefore, in the following analysis, we set the theoretical upper limit for the cross section calculations to $T$ = 150 MeV.

\begin{figure}[!b]
    \centering
    \includegraphics[width=1.0\linewidth]{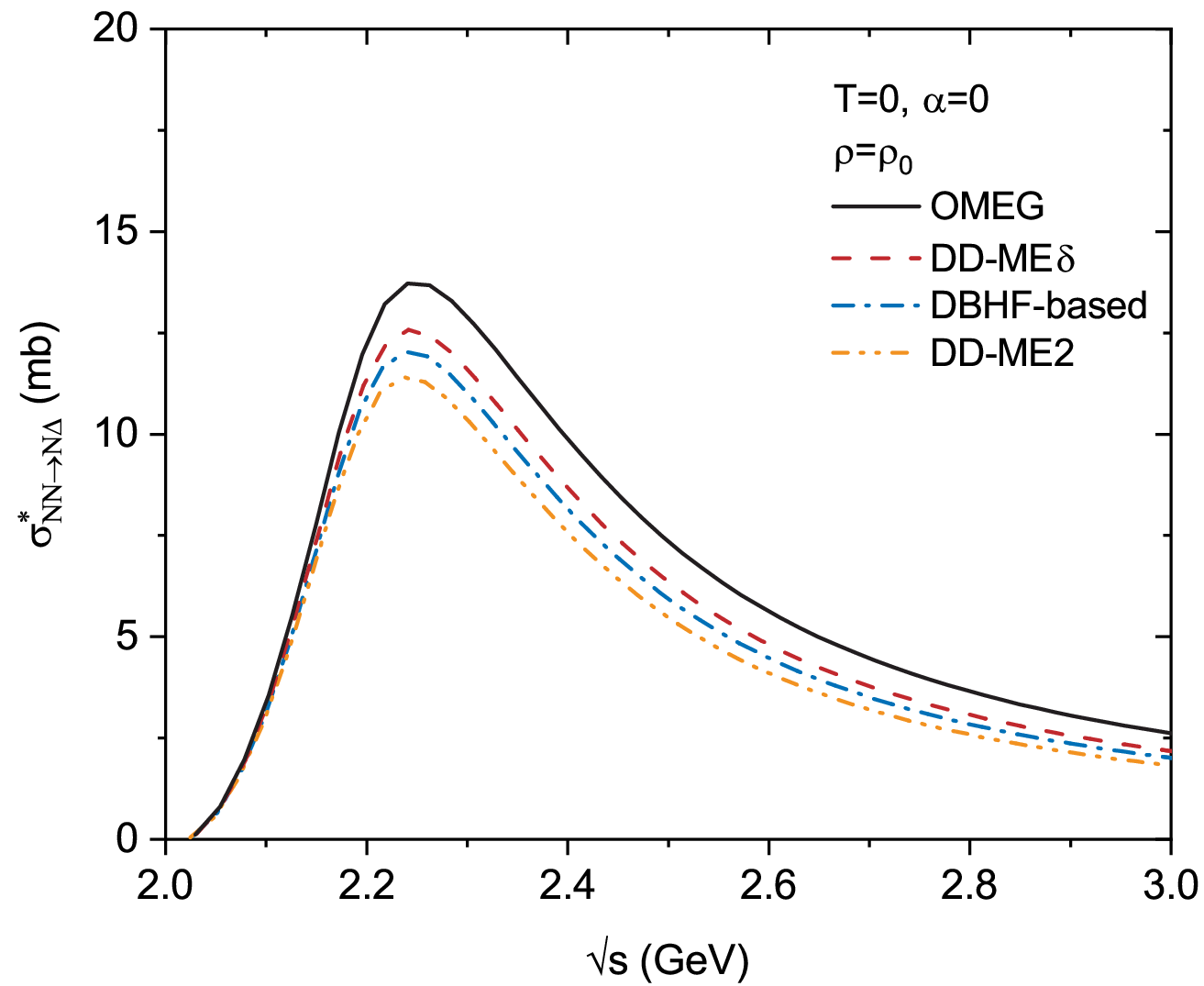}
   \caption{
   (Color online) The energy dependence of in-medium $NN$ inelastic cross section in isospin-symmetry ($\alpha=0$) nuclear medium with $T$=0 and $\rho=\rho_0$. The black solid, red dashed, blue dash-dotted, and orange dash-dot-dotted lines correspond to the results of OMEG, DD-ME$\delta$, DBHF-based, and DD-ME2 parametrization, respectively.}
    \label{fig.1.5}
\end{figure}

\begin{figure*}[t!]
    \centering
    \includegraphics[width=0.8\linewidth]{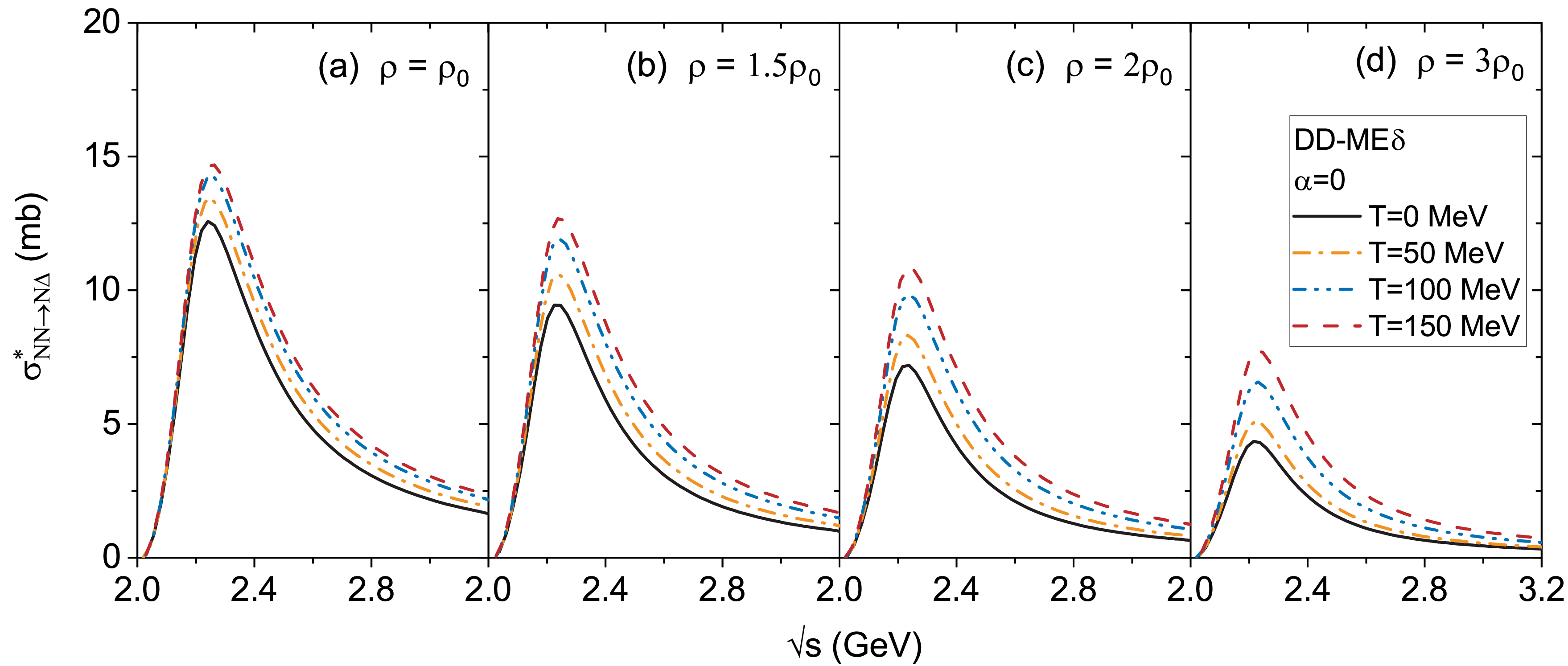}
   \caption{(Color online) The energy, density, and temperature dependence of the $\sigma^{*}_{NN\rightarrow N\Delta}$ in the isospin-symmetric nuclear matter calculated with the DD-ME$\delta$ parameter set.}
    \label{alpha=0 cross section DD-ME}
\end{figure*}

\begin{figure*}[htbp!]
    \centering
    \includegraphics[width=0.8\linewidth]{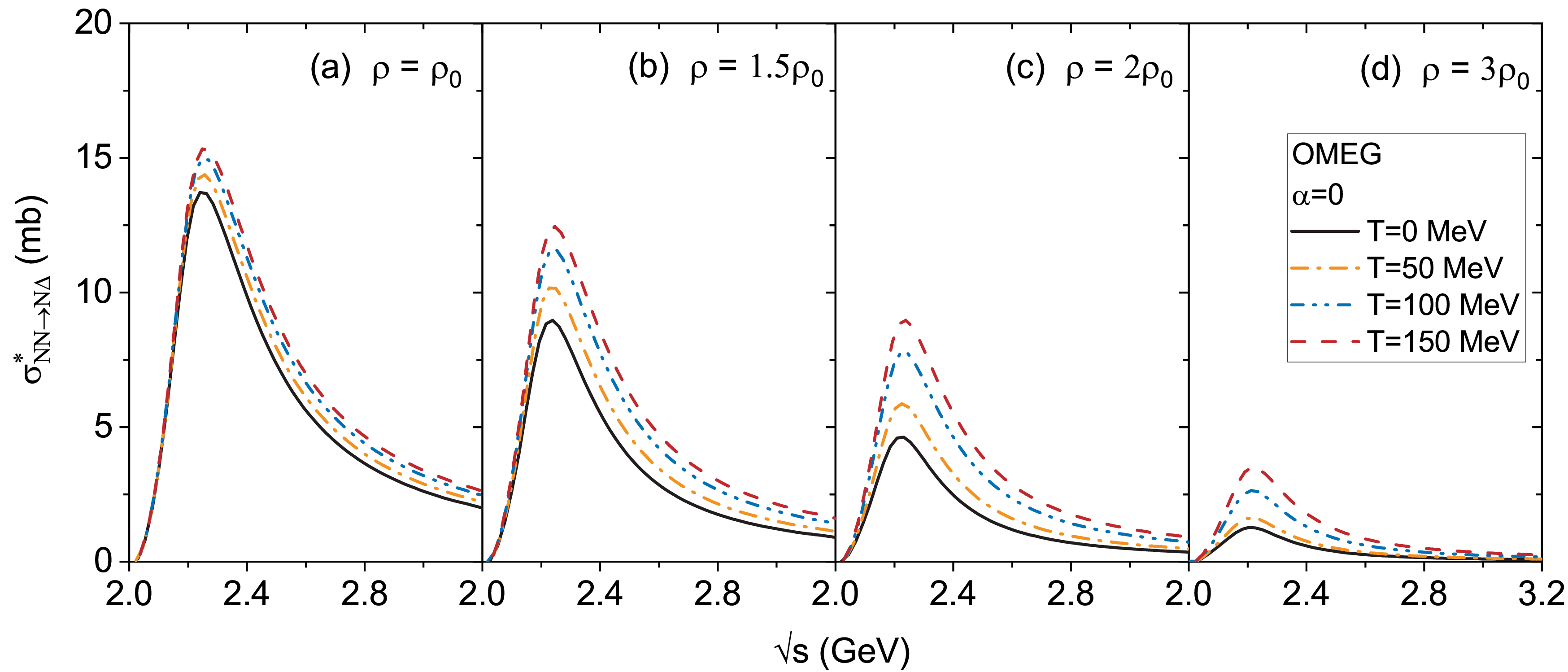}
   \caption{(Color online) Same as Fig. \ref{alpha=0 cross section DD-ME}, but for the results calculated with OMEG parameter set. }
    \label{alpha=0 cross section OMEG}
\end{figure*}

Considering that the $\Delta$ isobar is an unstable resonance in nature, its mass distribution in free space $m_{\Delta}$ is treated in the same way as in Ref. \cite{Mao:1994zza}, and the coupling of the $\Delta$ isobar to meson is assumed to be the same as that of nucleons. 
Therefore, the effective masses of four-charge $\Delta$ isobars, which are related to the nucleon effective mass by the Clebsch-Gordan coefficients of $\Delta \rightarrow N\pi$ decomposition, can be expressed as
\begin{equation}
\begin{array}{l}
m_{\Delta^{++} / \Delta^{-}}^{*}=m_{\Delta}-g^{\sigma}_{\Delta\Delta} \sigma \mp g^{\delta}_{\Delta\Delta} \delta_{}, \\
m_{\Delta^{+} / \Delta^{0}}^{*}=m_{\Delta}-g^{\sigma}_{\Delta\Delta} \sigma \mp \frac{1}{3} g^{\delta}_{\Delta\Delta} \delta_{}.
\label{delta mass}
\end{array}
\end{equation}
Since the effective masses of $\Delta$ isotopes exhibit similar density and temperature dependence as those of protons and neutrons shown in Fig.~\ref{fig.1}, they are not shown here.


After introducing the mean-field framework, we briefly summarize the main ingredients of the approach employed in this work. The scalar and vector meson fields are obtained by solving the corresponding field equations, which determine the Dirac effective masses of baryons in the medium. The analytical expressions for the spin and isospin matrices entering the collision terms are derived from the collision self-energies within the RBUU approach and were presented in Eqs. (31)–(38) of Ref.~\cite{Mao:1994zza}. In the calculation of the in-medium cross sections, the free baryon masses appearing in the spin matrices are replaced by the corresponding in-medium effective masses of the incoming and outgoing nucleons and $\Delta$ given in Eq. \ref{nucleon mass} and \ref{delta mass}.
For the nucleon–nucleon–$\pi$ and nucleon–$\Delta$–$\pi$ vertex, the same coupling as inferred in Ref.~\cite{Li:2016xix,Mao:1994zza} is adopted. Additionally, due to the finite size of hadrons and a part of the short-range correlation, the effective form factor of $NN\pi$ and $N\Delta\pi$ vertex is taken as in Ref. \cite{Mao:1994zza}.


It should be noted that HICs are intrinsically non-equilibrated systems \cite{Li:1993muy}. Consequently, the temperature discussed in experimental analyses and transport model studies is often introduced in an effective sense, for example through slope parameters extracted from particle spectra \cite{Gaitanos:2003zg,Reichert:2022qys,Reichert:2020uxs}. In the present work, the temperature appearing in the finite-temperature relativistic mean-field framework is likewise treated as an effective theoretical parameter, which is employed to explore how temperature-dependent in-medium properties of nuclear matter may influence dynamical quantities, such as $NN$ inelastic cross sections.

\section{Results and discussion}\label{sec3}

To investigate the influence of different approaches on the in-medium $NN$ inelastic cross sections, 
Fig.~\ref{fig.1.5} compares the results obtained with four representative interactions, namely DD-ME$\delta$, OMEG, DD-ME2 (as used in Ref.~\cite{Nan:2023gwp}), and the parametrization fitted to DBHF self-energies (as used in Ref.~\cite{Li:2003vd}), in isospin-symmetric nuclear matter at $T=0$ and $\rho=\rho_0$. 
It is observed that these results exhibit only a weak model dependence across the entire energy range considered. 
Among these parametrizations, OMEG yields the largest cross section, exceeding the smallest one (DD-ME2) by about 2~mb around the peak region. The results obtained with the DD-ME$\delta$ and DBHF-based parametrizations are very close to each other, indicating comparable effective masses at saturation density.

\subsection{Temperature effect on the $\sigma^{*}_{NN \rightarrow N\Delta}(\sqrt{s},\rho)$ in isospin-symmetric nuclear medium}

\begin{figure}[t]
    \centering
    \includegraphics[width=0.9\linewidth]{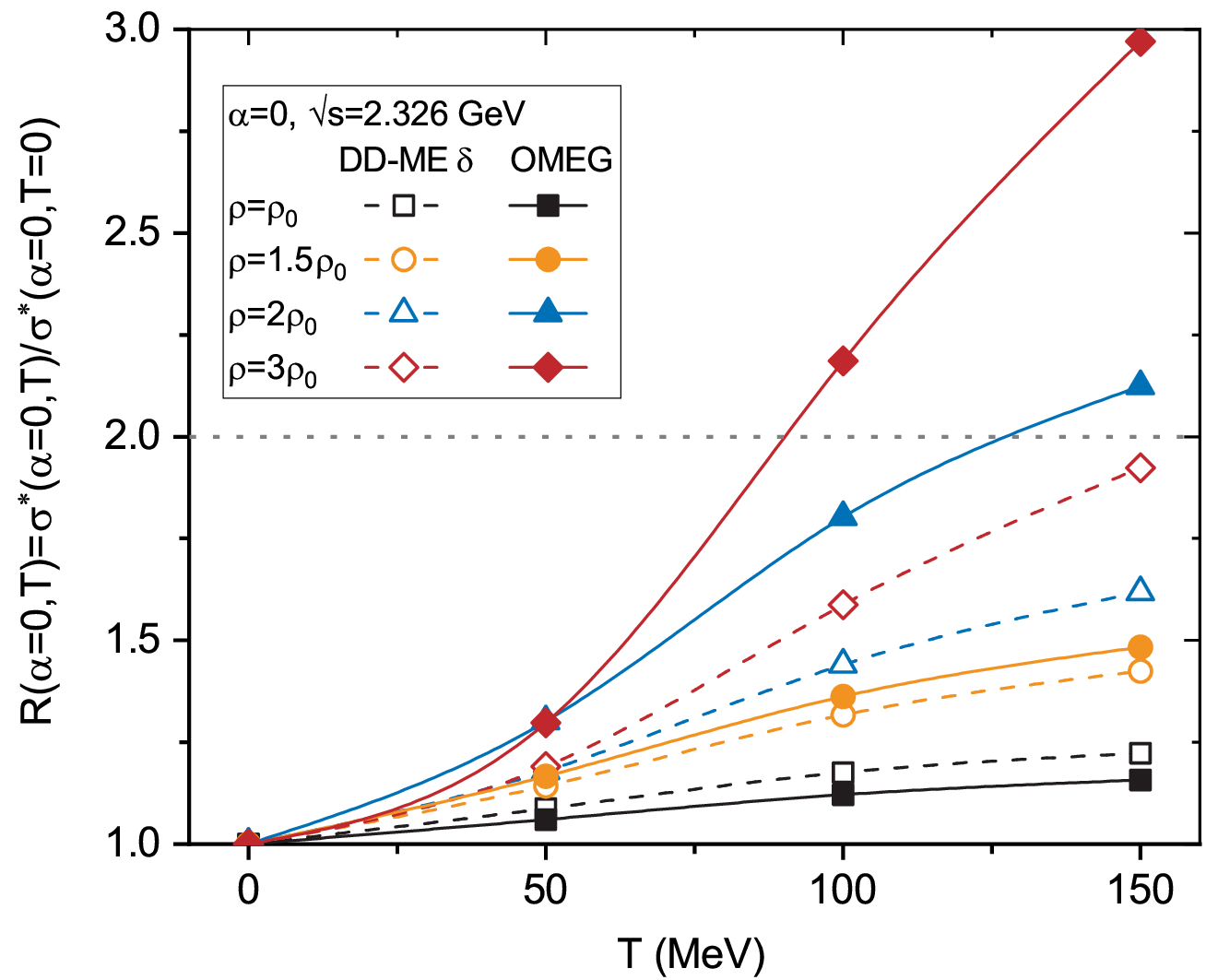}
   \caption{(Color online) Ratios $R(\alpha=0, T)=\sigma^{*}(\alpha = 0, T)/\sigma^{*}(\alpha = 0, T=0)$ for the $\Delta$ production channels as functions of temperature $T$ at $\sqrt{s} = 2.326$ GeV at densities of $\rho_{0}$, 1.5$\rho_{0}$, 2$\rho_{0}$, and 3$\rho_{0}$. The horizontal gray dotted line denotes the reference value of 2.}
    \label{alpha=0 ratio}
\end{figure}

First, the temperature effect on the energy- and density-dependent $NN \rightarrow N\Delta$ cross section $\sigma^{*}_{NN \rightarrow N\Delta}$ in the isospin-symmetric nuclear matter ($\alpha$=0) is investigated. 
To highlight the most pronounced isospin effects, only the $\Delta^{++}$ and $\Delta^{-}$ states are considered in this work. 
The calculated results using the DD-ME$\delta$ and OMEG parameter sets are presented in Fig. \ref{alpha=0 cross section DD-ME} and Fig. \ref{alpha=0 cross section OMEG}. From left to right, the results correspond to $\rho = \rho_0$, $1.5\rho_0$, $2\rho_0$, and $3\rho_0$, respectively. 
It is observed that the cross section increases steeply when the center-of-mass (c.m.) energy is near threshold ($\sqrt{s}\le 2.3$ GeV), followed by a gradual decrease at higher energies. This near-threshold enhancement originates from the phase-space factor in the cross-section formula, i.e., 
\begin{equation}
\sigma^*_{NN\rightarrow N\Delta} \propto \frac{|\vec{p}_{f}|}{|\vec{p}_{i}|}|M|^{2},
\end{equation}
where $|\vec{p}_i|$ and $|\vec{p}_f|$ are the c.m. momenta of the incoming and outgoing particles, respectively, and $|M|^{2}$ is the square of the transition matrix element. Above the threshold, $|\vec{p}_f|$ increases rapidly, causing the ratio $|\vec{p}_f|/|\vec{p}_i|$ to rise sharply, thereby amplifying the cross section.
As the density increases, the $\Delta$ production cross sections decrease monotonically, mainly due to the reduction of the effective baryon masses with increasing density. 
And the results for the OMEG set exhibit a more rapid decrease compared to those of the DD-ME$\delta$ case, due to the stronger density dependence of the effective mass, as shown in the left panels of Fig.~\ref{fig.1}. 
Further, the influence of finite temperature is examined by comparing results at $T=50$ (orange dash-dotted lines), 100 (blue dash-dot-dotted lines), and 150 MeV (red dashed lines) with those at $T=0$ MeV (black solid lines). 
As the temperature increases, the cross section is enhanced because the baryon effective masses increase with finite temperature. The similar temperature effect can also be seen in the nucleon–nucleon elastic cross sections $\sigma^{*}_{NN \rightarrow NN}$ \cite{Li:2003vd}. 
Furthermore, although the baryon effective masses of these two parameter sets exhibit similar temperature dependence and different density dependence, at lower densities ($\rho=\rho_0$ and $1.5\rho_0$), the temperature dependence of the $\sigma^{*}_{NN \rightarrow N\Delta}$ is nearly identical for the DD-ME$\delta$ and OMEG sets, while the OMEG results display a much stronger sensitivity to the temperature at higher densities. 

To better quantify the effect of finite temperature on the $NN$ inelastic cross section, the temperature dependence of the ratios $R(\alpha = 0,T)=\sigma^{*}(\alpha = 0,T)/\sigma^{*}(\alpha = 0,T=0)$ at various densities is shown in Fig.~\ref{alpha=0 ratio}. 
A typical kinetic energy $E_K$=1 GeV (corresponding to $\sqrt{s} = 2.326$ GeV) is chosen for calculation. 
The dashed lines with open symbols and solid lines with solid symbols correspond to the results for DD-ME$\delta$ and OMEG sets, while the black, orange, blue, and red lines with symbols represent the results at $\rho_0$, $1.5\rho_0$, $2\rho_0$, and $3\rho_0$, respectively. 
It is seen that the ratio $R(\alpha=0, T)$ increases as temperature increases from 0 to 150 MeV. 
At the density below approximately $1.5\rho_0$, the increases in the ratios are nearly identical for both coupling sets.
However, at higher densities ($\rho \ge 1.5\rho_0$), the increases for the OMEG set (solid lines with solid symbols) are larger than those for the DD-ME$\delta$ set (dashed lines with open symbols). 
These results indicate that the temperature dependence of the $NN$ inelastic cross section in isospin-symmetric nuclear matter will be enhanced with increasing density, and the temperature and density dependence are different for different mean-field modes, especially at higher densities.



\begin{figure}[t]
    \centering
    \includegraphics[width=0.9\linewidth]{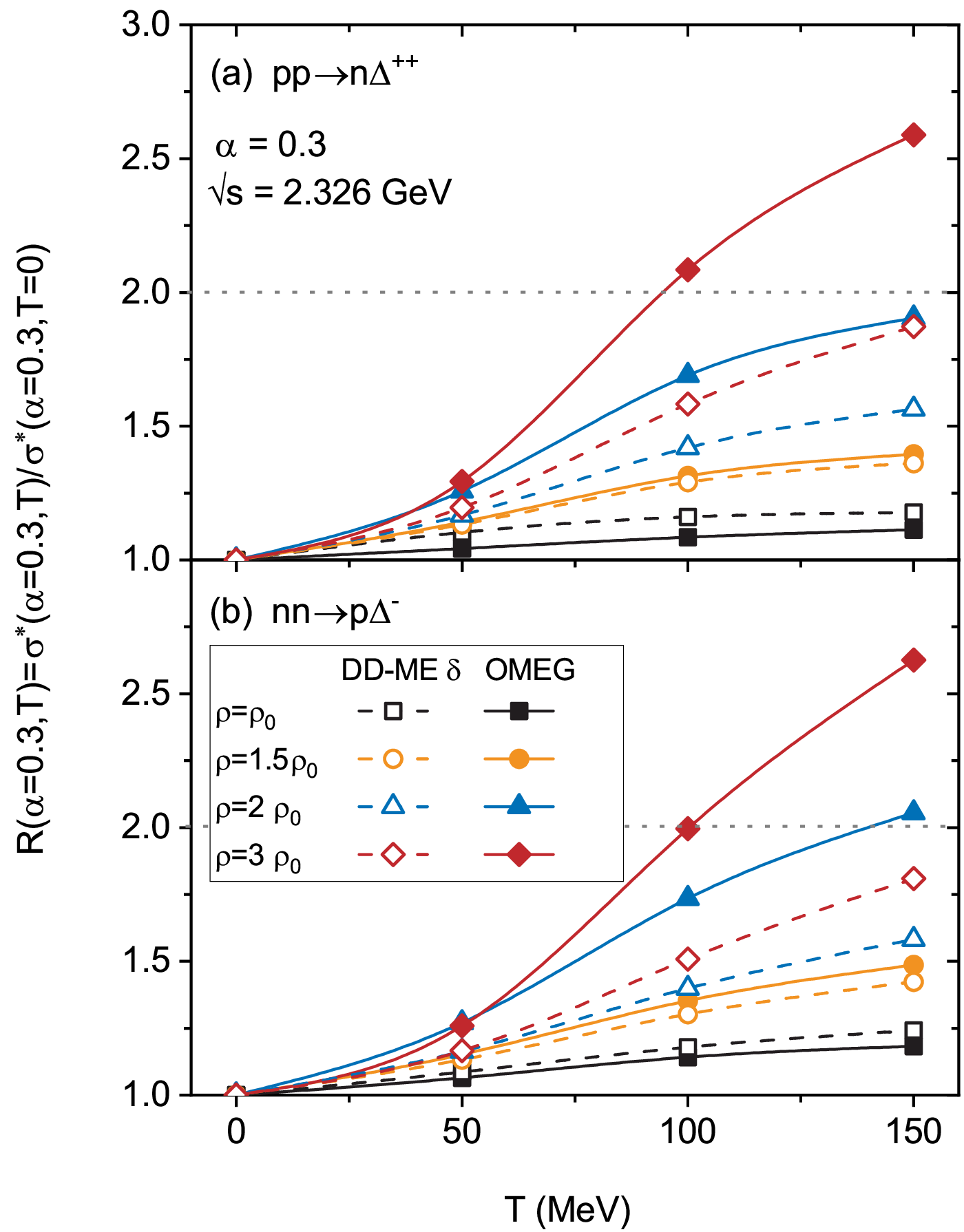}
   \caption{(Color online) The ratios $R(\alpha = 0.3,T)=\sigma^{*}(\alpha = 0.3,T)/\sigma^{*}(\alpha = 0.3,T=0)$ for $\Delta^{++}$ (top panel) and $\Delta^{-}$ (bottom panel) stiff production channels as a function of the temperature with the c.m. energy $\sqrt{s} = 2.326$ GeV at $\rho_{0}$, 1.5$\rho_{0}$, 2$\rho_{0}$, 3$\rho_{0}$. The horizontal gray dotted line denotes the reference value of 2.}
    \label{figure-cxt4}
\end{figure}

\begin{figure}[t]
    \centering
    \includegraphics[width=0.9\linewidth]{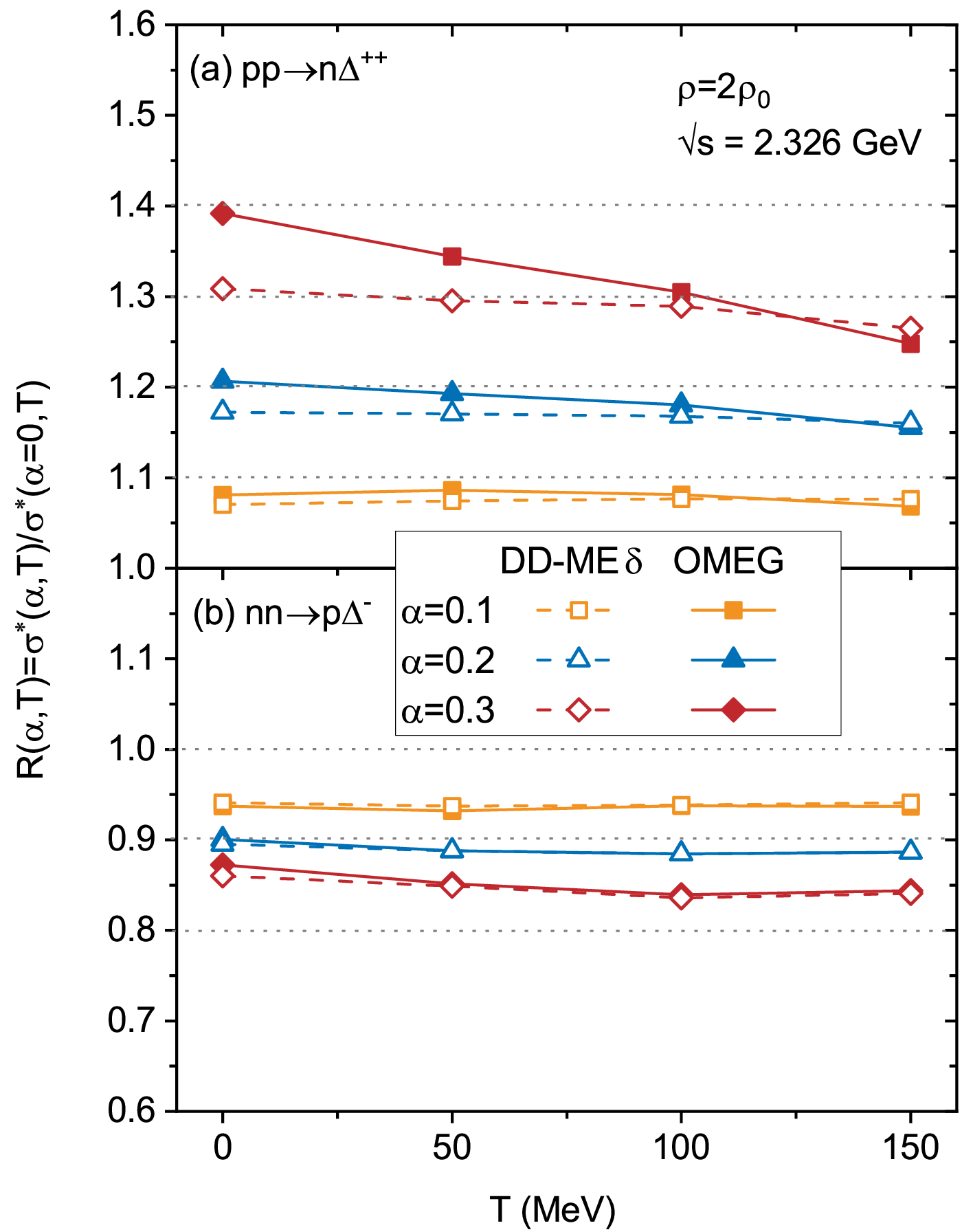}
   \caption{(Color online) Same as Fig. \ref{figure-cxt4}, but for $R(\alpha,T)=\sigma^{*}(\alpha,T)/\sigma^{*}(\alpha = 0.0,T)$ at 2$\rho_{0}$ with various $\alpha$.}
    \label{figure-5}
\end{figure}

\subsection{Temperature effect on the $\sigma^{*}_{NN \rightarrow N\Delta}(\sqrt{s},\rho)$ in isospin-asymmetric nuclear medium}

Further, the temperature effect on the $NN$ inelastic cross section in isospin-asymmetric nuclear matter is investigated.
In the isospin-asymmetric nuclear medium, the isospin matrices $T_{e,d}$ for both direct and exchanged Feynman diagrams, which contribute to the lowest order collisional self-energy for $\Delta$ production, are equal to 2 for the $\Delta^{++}$ and $\Delta^{-}$ production channels, and the $\delta$ meson exchange contributes opposite signs to the effective masses of $\Delta^{++}$ and $\Delta^{-}$, and of protons and neutrons, resulting in the splitting of $\sigma^{*}_{NN\rightarrow N\Delta}$ \cite{Li:2016xix,Mao:1994zza}. 
In addition, the energy, density, and temperature dependence of the $\Delta$ production cross sections in isospin-asymmetric nuclear matter are similar to those in isospin-symmetric nuclear matter. As the density increases, the cross section is suppressed, while an increase in temperature leads to a significant enhancement of the cross section. Therefore, this behavior does not need to be illustrated repeatedly for different isospin asymmetries.

Also, to better quantify the effect of finite temperature on the $NN$ inelastic cross section in isospin-asymmetry nuclear matter, the ratios $R(\alpha = 0.3,T)=\sigma^{*}(\alpha = 0.3,T)/ \\ \sigma^{*}(\alpha = 0.3,T=0)$ at $\sqrt{s}=2.326$ GeV are plotted in Fig.~\ref{figure-cxt4}. 
The upper and lower panels correspond to the ratios for $\sigma^{*}_{pp \rightarrow n\Delta^{++}}$ and $\sigma^{*}_{nn \rightarrow p\Delta^{-}}$, respectively. 
The similar density and temperature dependence of the cross section as shown in Fig. \ref{alpha=0 ratio} can be seen again. 
In addition, $R(\alpha = 0,T)\approx  R(\alpha = 0.3,T)$ at densities of $\rho=\rho_{0}$ and $1.5\rho_{0}$, while $R(\alpha = 0,T)> R(\alpha = 0.3,T)$ at higher densities. 
These results indicate that the temperature dependence of the $NN$ inelastic cross section in isospin-asymmetric nuclear matter will be suppressed compared to isospin-symmetric nuclear matter, especially at higher densities. 


\subsection{Temperature effect on the isospin dependence of the\\ 
 $\sigma^{*}_{NN \rightarrow N\Delta}(\sqrt{s},\rho,\alpha)$ in isospin-asymmetric nuclear \\ medium}

Last, the temperature effect on the isospin dependence of $\Delta$ production cross section is investigated. 
Taking $\rho = 2\rho_0$ as a representative case of moderately high density, Fig.~\ref{figure-5} shows the ratios $R(\alpha, T)$ for the $pp \rightarrow n\Delta^{++}$ and $nn \rightarrow p\Delta^{-}$ channels as functions of temperature $T$, with $\alpha$ = 0.1, 0.2, and 0.3 at $\sqrt{s}$ = 2.326 GeV. 
It is seen that $R(\alpha, T)$ exhibits a notable isospin dependence. 
As $\alpha$ increases from 0 to 0.3, $R(\alpha, T)$ for $pp \rightarrow n\Delta^{++}$ increases, whereas that for $nn \rightarrow p\Delta^{-}$ decreases. 
This behavior arises because the in-medium corrections to the effective masses of $\Delta^{++}$ and $\Delta^{-}$ have opposite signs when $\alpha \neq 0$ \cite{Li:2016xix}. 
Moreover, for the $nn \rightarrow p\Delta^{-}$ channel with relatively lower isospin asymmetry ($\alpha=0.1$ and/or 0.2), the ratios of the two methods are found to be independent of temperature, while they are weakly dependent on temperature at $\alpha=0.3$. 
As for $pp \rightarrow n\Delta^{++}$, the temperature effect on the isospin dependence of the ratio calculated with DD-ME$\delta$ is the same as that for $nn \rightarrow p\Delta^{-}$ channel. While the isospin dependence of the $\sigma^{*}_{pp \rightarrow n\Delta^{++}}$ calculated with OMEG is enhanced at low temperature, and shows a relative suppression at higher temperature. 
In other words, the temperature dependence of the $\sigma^{*}_{pp \rightarrow n\Delta^{++}}$ from calculations with OMEG becomes more pronounced as the isospin increases. 
These results indicate that the isospin dependence of the $NN$ inelastic cross section in the hot and isospin-asymmetric nuclear matter is almost temperature-independent at lower isospin asymmetries, while for a higher isospin-asymmetric system, the isospin dependence is not only influenced by the nucleon–meson coupling parameter set, but also by the temperature.

Since Ref.~\cite{Li:2016xix}, the in-medium production and absorption cross sections of the $\Delta$ resonance
($\sigma^{*}_{NN \rightarrow N\Delta}$, $\sigma^{*}_{N\Delta \rightarrow N\Delta}$, $\sigma^{*}_{N\pi \rightarrow \Delta}$)
have been systematically studied within the RBUU theoretical framework
\cite{Li:2016xix,Li:2017pis,Nan:2023gwp,Nan:2024ogc}.
By comparing the corresponding $R(\alpha, T)$ ratios among these three channels for the production of the same $\Delta$ isobar,
it is found that the in-medium effect on the $N\Delta \rightarrow N\Delta$ cross section tends to amplify
the isospin dependence of the $NN \rightarrow N\Delta$ cross section,
whereas the temperature effect in the $NN \rightarrow N\Delta$ channel and the isospin effect in the
$N\pi \rightarrow \Delta$ channel act to moderate the isospin dependence of $\Delta$ production.
To obtain a comprehensive understanding of the in-medium effects on $\Delta$ and pion observables,
and to derive reliable constraints on the EoS at high-baryon density through direct comparison with experimental data,
these energy-, density-, isospin-, and temperature-dependent in-medium cross sections
have not yet been incorporated into transport models.
At this stage, it is timely and well motivated to implement these in-medium cross sections into transport simulations,
however, such investigations are beyond the scope of the present work and will be pursued in forthcoming studies.


\section{Summary and outlook}\label{sec4}

To summarize, within the self-consistent relativistic  Boltzmann Uehling-Uhlenbeck (RBUU) 
framework, the temperature effects on the energy-, density-, and isospin-dependent $NN$ inelastic cross section, taking $pp \to n\Delta^{++}$ and $nn \to p\Delta^{-}$ as examples, are investigated by using two types of relativistic mean-field couplings, the density-dependent parameterization DD-ME$\delta$ and the nonlinear parameterization OMEG. 
It is found that: 
\begin{enumerate}[label=(\roman*)]
    \item For the isospin-symmetric and -asymmetric nuclear medium, as the temperature increases from 0 to 150 MeV, the $\Delta$ stiff production cross sections show a significant enhancement under both the DD-ME$\delta$ and OMEG sets. 
    These increases are nearly identical in both coupling sets at densities below about $1.5\rho_0$. At higher densities, however, the enhancement in the DD-ME$\delta$ case is smaller than that in the OMEG case. \\
    \item  As the density $\rho$ increases, the cross sections decrease, and the temperature dependence becomes more pronounced. Moreover, the results of the OMEG mode display a stronger density and temperature dependence compared to those of the DD-ME$\delta$ case, since the baryon effective masses in the OMEG mode have a much stronger density dependence. \\
    \item  For the isospin-asymmetric nuclear medium, the temperature dependence of the $NN$ inelastic cross section in isospin-asymmetric nuclear matter will be suppressed to some extent compared to the isospin-symmetric nuclear medium, especially at higher densities. \\
    \item  As the isospin asymmetry $\alpha$ varies from 0.1 to 0.3, the isospin dependence of the cross section in a hot isospin-asymmetric nuclear matter is almost independent of temperature at lower isospin asymmetries, while for a higher isospin-asymmetric system, the isospin dependence becomes complicated because of the model dependence.    \\
\end{enumerate} 

These results indicate that both the OMEG and DD-ME$\delta$ parameter sets, which are constrained by astrophysical observations and finite nuclei calculations, respectively, provide a self-consistent description of $\Delta$ production cross sections in HICs at densities below approximately $1.5\rho_{0}$. At higher densities, however, the two models exhibit distinct density-dependent behaviors. Combined with data from upcoming HIC experiments at HIAF, FAIR, and other facilities, this model dependence will be effectively reduced, providing important constraints on the equation of state at high density, which is crucial for understanding nuclear reactions and neutron star merger dynamics.

\section*{Acknowledgements}
The study is supported in part by the National Natural Science Foundation of China (Nos. 12335008 and 12505143), the National Key Research and Development Program of China under (Grant No. 2023YFA1606402), the National Natural Science Foundation of Zhejiang Province (No. LQN25A050003), the Huzhou Natural Science Foundation (2024YZ28), the Scientific Research Fund of the Zhejiang Provincial Education Department (No. Y202353782), and the Foundation of National Key Laboratory of Plasma Physics (Grant No. 6142A04230203). The authors are grateful to the C3S2 computing center in Huzhou University for calculation support.

\bibliography{ref.bib}

\begin{thebibliography}{69}%
\makeatletter
\providecommand \@ifxundefined [1]{%
 \@ifx{#1\undefined}
}%
\providecommand \@ifnum [1]{%
 \ifnum #1\expandafter \@firstoftwo
 \else \expandafter \@secondoftwo
 \fi
}%
\providecommand \@ifx [1]{%
 \ifx #1\expandafter \@firstoftwo
 \else \expandafter \@secondoftwo
 \fi
}%
\providecommand \natexlab [1]{#1}%
\providecommand \enquote  [1]{``#1''}%
\providecommand \bibnamefont  [1]{#1}%
\providecommand \bibfnamefont [1]{#1}%
\providecommand \citenamefont [1]{#1}%
\providecommand \href@noop [0]{\@secondoftwo}%
\providecommand \href [0]{\begingroup \@sanitize@url \@href}%
\providecommand \@href[1]{\@@startlink{#1}\@@href}%
\providecommand \@@href[1]{\endgroup#1\@@endlink}%
\providecommand \@sanitize@url [0]{\catcode `\\12\catcode `\$12\catcode
  `\&12\catcode `\#12\catcode `\^12\catcode `\_12\catcode `\%12\relax}%
\providecommand \@@startlink[1]{}%
\providecommand \@@endlink[0]{}%
\providecommand \url  [0]{\begingroup\@sanitize@url \@url }%
\providecommand \@url [1]{\endgroup\@href {#1}{\urlprefix }}%
\providecommand \urlprefix  [0]{URL }%
\providecommand \Eprint [0]{\href }%
\providecommand \doibase [0]{http://dx.doi.org/}%
\providecommand \selectlanguage [0]{\@gobble}%
\providecommand \bibinfo  [0]{\@secondoftwo}%
\providecommand \bibfield  [0]{\@secondoftwo}%
\providecommand \translation [1]{[#1]}%
\providecommand \BibitemOpen [0]{}%
\providecommand \bibitemStop [0]{}%
\providecommand \bibitemNoStop [0]{.\EOS\space}%
\providecommand \EOS [0]{\spacefactor3000\relax}%
\providecommand \BibitemShut  [1]{\csname bibitem#1\endcsname}%
\let\auto@bib@innerbib\@empty
\bibitem [{\citenamefont {Teis}\ \emph {et~al.}(1997)\citenamefont {Teis},
  \citenamefont {Cassing}, \citenamefont {Effenberger}, \citenamefont
  {Hombach}, \citenamefont {Mosel},\ and\ \citenamefont {Wolf}}]{Teis:1996kx}%
  \BibitemOpen
  \bibfield  {author} {\bibinfo {author} {\bibfnamefont {S.}~\bibnamefont
  {Teis}}, \bibinfo {author} {\bibfnamefont {W.}~\bibnamefont {Cassing}},
  \bibinfo {author} {\bibfnamefont {M.}~\bibnamefont {Effenberger}}, \bibinfo
  {author} {\bibfnamefont {A.}~\bibnamefont {Hombach}}, \bibinfo {author}
  {\bibfnamefont {U.}~\bibnamefont {Mosel}}, \ and\ \bibinfo {author}
  {\bibfnamefont {G.}~\bibnamefont {Wolf}},\ }\href {\doibase
  10.1007/s002180050198} {\bibfield  {journal} {\bibinfo  {journal} {Z. Phys.
  A}\ }\textbf {\bibinfo {volume} {356}},\ \bibinfo {pages} {421} (\bibinfo
  {year} {1997})},\ \Eprint {http://arxiv.org/abs/nucl-th/9609009}
  {arXiv:nucl-th/9609009} \BibitemShut {NoStop}%
\bibitem [{\citenamefont {Li}\ and\ \citenamefont {Ko}(1995)}]{Li:1995pra}%
  \BibitemOpen
  \bibfield  {author} {\bibinfo {author} {\bibfnamefont {B.~A.}\ \bibnamefont
  {Li}}\ and\ \bibinfo {author} {\bibfnamefont {C.~M.}\ \bibnamefont {Ko}},\
  }\href {\doibase 10.1103/PhysRevC.52.2037} {\bibfield  {journal} {\bibinfo
  {journal} {Phys. Rev. C}\ }\textbf {\bibinfo {volume} {52}},\ \bibinfo
  {pages} {2037} (\bibinfo {year} {1995})},\ \Eprint
  {http://arxiv.org/abs/nucl-th/9505016} {arXiv:nucl-th/9505016} \BibitemShut
  {NoStop}%
\bibitem [{\citenamefont {Stoecker}\ and\ \citenamefont
  {Greiner}(1986)}]{Stoecker:1986ci}%
  \BibitemOpen
  \bibfield  {author} {\bibinfo {author} {\bibfnamefont {H.}~\bibnamefont
  {Stoecker}}\ and\ \bibinfo {author} {\bibfnamefont {W.}~\bibnamefont
  {Greiner}},\ }\href {\doibase 10.1016/0370-1573(86)90131-6} {\bibfield
  {journal} {\bibinfo  {journal} {Phys. Rept.}\ }\textbf {\bibinfo {volume}
  {137}},\ \bibinfo {pages} {277} (\bibinfo {year} {1986})}\BibitemShut
  {NoStop}%
\bibitem [{\citenamefont {Cai}\ \emph {et~al.}(2015)\citenamefont {Cai},
  \citenamefont {Fattoyev}, \citenamefont {Li},\ and\ \citenamefont
  {Newton}}]{Cai:2015hya}%
  \BibitemOpen
  \bibfield  {author} {\bibinfo {author} {\bibfnamefont {B.~J.}\ \bibnamefont
  {Cai}}, \bibinfo {author} {\bibfnamefont {F.~J.}\ \bibnamefont {Fattoyev}},
  \bibinfo {author} {\bibfnamefont {B.~A.}\ \bibnamefont {Li}}, \ and\ \bibinfo
  {author} {\bibfnamefont {W.~G.}\ \bibnamefont {Newton}},\ }\href {\doibase
  10.1103/PhysRevC.92.015802} {\bibfield  {journal} {\bibinfo  {journal} {Phys.
  Rev. C}\ }\textbf {\bibinfo {volume} {92}},\ \bibinfo {pages} {015802}
  (\bibinfo {year} {2015})},\ \Eprint {http://arxiv.org/abs/1501.01680}
  {arXiv:1501.01680 [nucl-th]} \BibitemShut {NoStop}%
\bibitem [{\citenamefont {Drago}\ \emph {et~al.}(2014)\citenamefont {Drago},
  \citenamefont {Lavagno}, \citenamefont {Pagliara},\ and\ \citenamefont
  {Pigato}}]{Drago:2014oja}%
  \BibitemOpen
  \bibfield  {author} {\bibinfo {author} {\bibfnamefont {A.}~\bibnamefont
  {Drago}}, \bibinfo {author} {\bibfnamefont {A.}~\bibnamefont {Lavagno}},
  \bibinfo {author} {\bibfnamefont {G.}~\bibnamefont {Pagliara}}, \ and\
  \bibinfo {author} {\bibfnamefont {D.}~\bibnamefont {Pigato}},\ }\href
  {\doibase 10.1103/PhysRevC.90.065809} {\bibfield  {journal} {\bibinfo
  {journal} {Phys. Rev. C}\ }\textbf {\bibinfo {volume} {90}},\ \bibinfo
  {pages} {065809} (\bibinfo {year} {2014})},\ \Eprint
  {http://arxiv.org/abs/1407.2843} {arXiv:1407.2843 [astro-ph.SR]} \BibitemShut
  {NoStop}%
\bibitem [{\citenamefont {Huber}\ and\ \citenamefont
  {Aichelin}(1994)}]{Huber:1994ee}%
  \BibitemOpen
  \bibfield  {author} {\bibinfo {author} {\bibfnamefont {S.}~\bibnamefont
  {Huber}}\ and\ \bibinfo {author} {\bibfnamefont {J.}~\bibnamefont
  {Aichelin}},\ }\href {\doibase 10.1016/0375-9474(94)90232-1} {\bibfield
  {journal} {\bibinfo  {journal} {Nucl. Phys. A}\ }\textbf {\bibinfo {volume}
  {573}},\ \bibinfo {pages} {587} (\bibinfo {year} {1994})}\BibitemShut
  {NoStop}%
\bibitem [{\citenamefont {Ter~Haar}\ and\ \citenamefont
  {Malfliet}(1987)}]{TerHaar:1986xpv}%
  \BibitemOpen
  \bibfield  {author} {\bibinfo {author} {\bibfnamefont {B.}~\bibnamefont
  {Ter~Haar}}\ and\ \bibinfo {author} {\bibfnamefont {R.}~\bibnamefont
  {Malfliet}},\ }\href {\doibase 10.1016/0370-1573(87)90085-8} {\bibfield
  {journal} {\bibinfo  {journal} {Phys. Rept.}\ }\textbf {\bibinfo {volume}
  {149}},\ \bibinfo {pages} {207} (\bibinfo {year} {1987})}\BibitemShut
  {NoStop}%
\bibitem [{\citenamefont {Mao}\ \emph {et~al.}(1994{\natexlab{a}})\citenamefont
  {Mao}, \citenamefont {Li}, \citenamefont {Zhuo}, \citenamefont {Han},\ and\
  \citenamefont {Yu}}]{Mao:1994zza}%
  \BibitemOpen
  \bibfield  {author} {\bibinfo {author} {\bibfnamefont {G.~J.}\ \bibnamefont
  {Mao}}, \bibinfo {author} {\bibfnamefont {Z.~X.}\ \bibnamefont {Li}},
  \bibinfo {author} {\bibfnamefont {Y.~Z.}\ \bibnamefont {Zhuo}}, \bibinfo
  {author} {\bibfnamefont {Y.~L.}\ \bibnamefont {Han}}, \ and\ \bibinfo
  {author} {\bibfnamefont {Z.~Q.}\ \bibnamefont {Yu}},\ }\href {\doibase
  10.1103/PhysRevC.49.3137} {\bibfield  {journal} {\bibinfo  {journal} {Phys.
  Rev. C}\ }\textbf {\bibinfo {volume} {49}},\ \bibinfo {pages} {3137}
  (\bibinfo {year} {1994}{\natexlab{a}})}\BibitemShut {NoStop}%
\bibitem [{\citenamefont {Li}\ \emph {et~al.}(2006)\citenamefont {Li},
  \citenamefont {Li}, \citenamefont {Soff}, \citenamefont {Bleicher},\ and\
  \citenamefont {Stoecker}}]{Li:2005gfa}%
  \BibitemOpen
  \bibfield  {author} {\bibinfo {author} {\bibfnamefont {Q.~F.}\ \bibnamefont
  {Li}}, \bibinfo {author} {\bibfnamefont {Z.~X.}\ \bibnamefont {Li}}, \bibinfo
  {author} {\bibfnamefont {S.}~\bibnamefont {Soff}}, \bibinfo {author}
  {\bibfnamefont {M.}~\bibnamefont {Bleicher}}, \ and\ \bibinfo {author}
  {\bibfnamefont {H.}~\bibnamefont {Stoecker}},\ }\href {\doibase
  10.1088/0954-3899/32/2/007} {\bibfield  {journal} {\bibinfo  {journal} {J.
  Phys. G}\ }\textbf {\bibinfo {volume} {32}},\ \bibinfo {pages} {151}
  (\bibinfo {year} {2006})},\ \Eprint {http://arxiv.org/abs/nucl-th/0509070}
  {arXiv:nucl-th/0509070} \BibitemShut {NoStop}%
\bibitem [{\citenamefont {Xiao}\ \emph {et~al.}(2009)\citenamefont {Xiao},
  \citenamefont {Li}, \citenamefont {Chen}, \citenamefont {Yong},\ and\
  \citenamefont {Zhang}}]{Xiao:2008vm}%
  \BibitemOpen
  \bibfield  {author} {\bibinfo {author} {\bibfnamefont {Z.~G.}\ \bibnamefont
  {Xiao}}, \bibinfo {author} {\bibfnamefont {B.~A.}\ \bibnamefont {Li}},
  \bibinfo {author} {\bibfnamefont {L.~W.}\ \bibnamefont {Chen}}, \bibinfo
  {author} {\bibfnamefont {G.~C.}\ \bibnamefont {Yong}}, \ and\ \bibinfo
  {author} {\bibfnamefont {M.}~\bibnamefont {Zhang}},\ }\href {\doibase
  10.1103/PhysRevLett.102.062502} {\bibfield  {journal} {\bibinfo  {journal}
  {Phys. Rev. Lett.}\ }\textbf {\bibinfo {volume} {102}},\ \bibinfo {pages}
  {062502} (\bibinfo {year} {2009})},\ \Eprint {http://arxiv.org/abs/0808.0186}
  {arXiv:0808.0186 [nucl-th]} \BibitemShut {NoStop}%
\bibitem [{\citenamefont {Ono}\ \emph {et~al.}(2019)\citenamefont {Ono} \emph
  {et~al.}}]{TMEP:2019yci}%
  \BibitemOpen
  \bibfield  {author} {\bibinfo {author} {\bibfnamefont {A.}~\bibnamefont
  {Ono}} \emph {et~al.} (\bibinfo {collaboration} {TMEP}),\ }\href {\doibase
  10.1103/PhysRevC.100.044617} {\bibfield  {journal} {\bibinfo  {journal}
  {Phys. Rev. C}\ }\textbf {\bibinfo {volume} {100}},\ \bibinfo {pages}
  {044617} (\bibinfo {year} {2019})},\ \Eprint
  {http://arxiv.org/abs/1904.02888} {arXiv:1904.02888 [nucl-th]} \BibitemShut
  {NoStop}%
\bibitem [{\citenamefont {Xu}\ \emph {et~al.}(2024)\citenamefont {Xu} \emph
  {et~al.}}]{TMEP:2023ifw}%
  \BibitemOpen
  \bibfield  {author} {\bibinfo {author} {\bibfnamefont {J.}~\bibnamefont {Xu}}
  \emph {et~al.} (\bibinfo {collaboration} {TMEP}),\ }\href {\doibase
  10.1103/PhysRevC.109.044609} {\bibfield  {journal} {\bibinfo  {journal}
  {Phys. Rev. C}\ }\textbf {\bibinfo {volume} {109}},\ \bibinfo {pages}
  {044609} (\bibinfo {year} {2024})},\ \Eprint
  {http://arxiv.org/abs/2308.05347} {arXiv:2308.05347 [nucl-th]} \BibitemShut
  {NoStop}%
\bibitem [{\citenamefont {Glendenning}(1985)}]{Glendenning:1984jr}%
  \BibitemOpen
  \bibfield  {author} {\bibinfo {author} {\bibfnamefont {N.~K.}\ \bibnamefont
  {Glendenning}},\ }\href {\doibase 10.1086/163253} {\bibfield  {journal}
  {\bibinfo  {journal} {Astrophys. J.}\ }\textbf {\bibinfo {volume} {293}},\
  \bibinfo {pages} {470} (\bibinfo {year} {1985})}\BibitemShut {NoStop}%
\bibitem [{\citenamefont {Glendenning}\ and\ \citenamefont
  {Schaffner-Bielich}(1998)}]{Glendenning:1998zx}%
  \BibitemOpen
  \bibfield  {author} {\bibinfo {author} {\bibfnamefont {N.~K.}\ \bibnamefont
  {Glendenning}}\ and\ \bibinfo {author} {\bibfnamefont {J.}~\bibnamefont
  {Schaffner-Bielich}},\ }\href {\doibase 10.1103/PhysRevLett.81.4564}
  {\bibfield  {journal} {\bibinfo  {journal} {Phys. Rev. Lett.}\ }\textbf
  {\bibinfo {volume} {81}},\ \bibinfo {pages} {4564} (\bibinfo {year}
  {1998})},\ \Eprint {http://arxiv.org/abs/astro-ph/9810284}
  {arXiv:astro-ph/9810284} \BibitemShut {NoStop}%
\bibitem [{\citenamefont {Sedrakian}\ and\ \citenamefont
  {Harutyunyan}(2022)}]{Sedrakian:2022kgj}%
  \BibitemOpen
  \bibfield  {author} {\bibinfo {author} {\bibfnamefont {A.}~\bibnamefont
  {Sedrakian}}\ and\ \bibinfo {author} {\bibfnamefont {A.}~\bibnamefont
  {Harutyunyan}},\ }\href {\doibase 10.1140/epja/s10050-022-00792-w} {\bibfield
   {journal} {\bibinfo  {journal} {Eur. Phys. J. A}\ }\textbf {\bibinfo
  {volume} {58}},\ \bibinfo {pages} {137} (\bibinfo {year} {2022})},\ \Eprint
  {http://arxiv.org/abs/2202.12083} {arXiv:2202.12083 [nucl-th]} \BibitemShut
  {NoStop}%
\bibitem [{\citenamefont {Li}(2002)}]{Li:2002qx}%
  \BibitemOpen
  \bibfield  {author} {\bibinfo {author} {\bibfnamefont {B.~A.}\ \bibnamefont
  {Li}},\ }\href {\doibase 10.1103/PhysRevLett.88.192701} {\bibfield  {journal}
  {\bibinfo  {journal} {Phys. Rev. Lett.}\ }\textbf {\bibinfo {volume} {88}},\
  \bibinfo {pages} {192701} (\bibinfo {year} {2002})},\ \Eprint
  {http://arxiv.org/abs/nucl-th/0205002} {arXiv:nucl-th/0205002} \BibitemShut
  {NoStop}%
\bibitem [{\citenamefont {Chen}\ \emph {et~al.}(2005)\citenamefont {Chen},
  \citenamefont {Ko},\ and\ \citenamefont {Li}}]{Chen:2004si}%
  \BibitemOpen
  \bibfield  {author} {\bibinfo {author} {\bibfnamefont {L.~W.}\ \bibnamefont
  {Chen}}, \bibinfo {author} {\bibfnamefont {C.~M.}\ \bibnamefont {Ko}}, \ and\
  \bibinfo {author} {\bibfnamefont {B.~A.}\ \bibnamefont {Li}},\ }\href
  {\doibase 10.1103/PhysRevLett.94.032701} {\bibfield  {journal} {\bibinfo
  {journal} {Phys. Rev. Lett.}\ }\textbf {\bibinfo {volume} {94}},\ \bibinfo
  {pages} {032701} (\bibinfo {year} {2005})},\ \Eprint
  {http://arxiv.org/abs/nucl-th/0407032} {arXiv:nucl-th/0407032} \BibitemShut
  {NoStop}%
\bibitem [{\citenamefont {Reisdorf}\ and\ \citenamefont {\textit{et
  al.}}(2007)}]{FOPI:2006ifg}%
  \BibitemOpen
  \bibfield  {author} {\bibinfo {author} {\bibfnamefont {W.}~\bibnamefont
  {Reisdorf}}\ and\ \bibinfo {author} {\bibnamefont {\textit{et al.}}}
  (\bibinfo {collaboration} {FOPI Collaboration}),\ }\href {\doibase
  10.1016/j.nuclphysa.2006.10.085} {\bibfield  {journal} {\bibinfo  {journal}
  {Nucl. Phys. A}\ }\textbf {\bibinfo {volume} {781}},\ \bibinfo {pages} {459}
  (\bibinfo {year} {2007})},\ \Eprint {http://arxiv.org/abs/nucl-ex/0610025}
  {arXiv:nucl-ex/0610025} \BibitemShut {NoStop}%
\bibitem [{\citenamefont {Jhang}\ and\ \citenamefont {\textit{et
  al.}}(2021)}]{SpiRIT:2020sfn}%
  \BibitemOpen
  \bibfield  {author} {\bibinfo {author} {\bibfnamefont {G.}~\bibnamefont
  {Jhang}}\ and\ \bibinfo {author} {\bibnamefont {\textit{et al.}}} (\bibinfo
  {collaboration} {SpiRIT Collaboration, TMEP Collaboration}),\ }\href
  {\doibase 10.1016/j.physletb.2020.136016} {\bibfield  {journal} {\bibinfo
  {journal} {Phys. Lett. B}\ }\textbf {\bibinfo {volume} {813}},\ \bibinfo
  {pages} {136016} (\bibinfo {year} {2021})},\ \Eprint
  {http://arxiv.org/abs/2012.06976} {arXiv:2012.06976 [nucl-ex]} \BibitemShut
  {NoStop}%
\bibitem [{\citenamefont {Guo}\ \emph {et~al.}(2014)\citenamefont {Guo},
  \citenamefont {Yong}, \citenamefont {Wang}, \citenamefont {Li}, \citenamefont
  {Zhang},\ and\ \citenamefont {Zuo}}]{Guo:2014tua}%
  \BibitemOpen
  \bibfield  {author} {\bibinfo {author} {\bibfnamefont {W.~M.}\ \bibnamefont
  {Guo}}, \bibinfo {author} {\bibfnamefont {G.~C.}\ \bibnamefont {Yong}},
  \bibinfo {author} {\bibfnamefont {Y.~J.}\ \bibnamefont {Wang}}, \bibinfo
  {author} {\bibfnamefont {Q.~F.}\ \bibnamefont {Li}}, \bibinfo {author}
  {\bibfnamefont {H.~F.}\ \bibnamefont {Zhang}}, \ and\ \bibinfo {author}
  {\bibfnamefont {W.}~\bibnamefont {Zuo}},\ }\href {\doibase
  10.1016/j.physletb.2014.10.011} {\bibfield  {journal} {\bibinfo  {journal}
  {Phys. Lett. B}\ }\textbf {\bibinfo {volume} {738}},\ \bibinfo {pages} {397}
  (\bibinfo {year} {2014})},\ \Eprint {http://arxiv.org/abs/1404.7217}
  {arXiv:1404.7217 [nucl-th]} \BibitemShut {NoStop}%
\bibitem [{\citenamefont {Li}\ \emph {et~al.}(2019)\citenamefont {Li},
  \citenamefont {Krastev}, \citenamefont {Wen},\ and\ \citenamefont
  {Zhang}}]{Li:2019xxz}%
  \BibitemOpen
  \bibfield  {author} {\bibinfo {author} {\bibfnamefont {B.~A.}\ \bibnamefont
  {Li}}, \bibinfo {author} {\bibfnamefont {P.~G.}\ \bibnamefont {Krastev}},
  \bibinfo {author} {\bibfnamefont {D.~H.}\ \bibnamefont {Wen}}, \ and\
  \bibinfo {author} {\bibfnamefont {N.~B.}\ \bibnamefont {Zhang}},\ }\href
  {\doibase 10.1140/epja/i2019-12780-8} {\bibfield  {journal} {\bibinfo
  {journal} {Eur. Phys. J. A}\ }\textbf {\bibinfo {volume} {55}},\ \bibinfo
  {pages} {117} (\bibinfo {year} {2019})},\ \Eprint
  {http://arxiv.org/abs/1905.13175} {arXiv:1905.13175 [nucl-th]} \BibitemShut
  {NoStop}%
\bibitem [{\citenamefont {Wolter}\ \emph {et~al.}(2022)\citenamefont {Wolter}
  \emph {et~al.}}]{TMEP:2022xjg}%
  \BibitemOpen
  \bibfield  {author} {\bibinfo {author} {\bibfnamefont {H.}~\bibnamefont
  {Wolter}} \emph {et~al.} (\bibinfo {collaboration} {TMEP}),\ }\href {\doibase
  10.1016/j.ppnp.2022.103962} {\bibfield  {journal} {\bibinfo  {journal} {Prog.
  Part. Nucl. Phys.}\ }\textbf {\bibinfo {volume} {125}},\ \bibinfo {pages}
  {103962} (\bibinfo {year} {2022})},\ \Eprint
  {http://arxiv.org/abs/2202.06672} {arXiv:2202.06672 [nucl-th]} \BibitemShut
  {NoStop}%
\bibitem [{\citenamefont {Bass}\ \emph {et~al.}(1995)\citenamefont {Bass},
  \citenamefont {Hartnack}, \citenamefont {Stoecker},\ and\ \citenamefont
  {Greiner}}]{Bass:1995pj}%
  \BibitemOpen
  \bibfield  {author} {\bibinfo {author} {\bibfnamefont {S.~A.}\ \bibnamefont
  {Bass}}, \bibinfo {author} {\bibfnamefont {C.}~\bibnamefont {Hartnack}},
  \bibinfo {author} {\bibfnamefont {H.}~\bibnamefont {Stoecker}}, \ and\
  \bibinfo {author} {\bibfnamefont {W.}~\bibnamefont {Greiner}},\ }\href
  {\doibase 10.1103/PhysRevC.51.3343} {\bibfield  {journal} {\bibinfo
  {journal} {Phys. Rev. C}\ }\textbf {\bibinfo {volume} {51}},\ \bibinfo
  {pages} {3343} (\bibinfo {year} {1995})},\ \Eprint
  {http://arxiv.org/abs/nucl-th/9501002} {arXiv:nucl-th/9501002} \BibitemShut
  {NoStop}%
\bibitem [{\citenamefont {Larionov}\ \emph {et~al.}(2001)\citenamefont
  {Larionov}, \citenamefont {Cassing}, \citenamefont {Leupold},\ and\
  \citenamefont {Mosel}}]{Larionov:2001va}%
  \BibitemOpen
  \bibfield  {author} {\bibinfo {author} {\bibfnamefont {A.~B.}\ \bibnamefont
  {Larionov}}, \bibinfo {author} {\bibfnamefont {W.}~\bibnamefont {Cassing}},
  \bibinfo {author} {\bibfnamefont {S.}~\bibnamefont {Leupold}}, \ and\
  \bibinfo {author} {\bibfnamefont {U.}~\bibnamefont {Mosel}},\ }\href
  {\doibase 10.1016/S0375-9474(01)01216-7} {\bibfield  {journal} {\bibinfo
  {journal} {Nucl. Phys. A}\ }\textbf {\bibinfo {volume} {696}},\ \bibinfo
  {pages} {747} (\bibinfo {year} {2001})},\ \Eprint
  {http://arxiv.org/abs/nucl-th/0103019} {arXiv:nucl-th/0103019} \BibitemShut
  {NoStop}%
\bibitem [{\citenamefont {Larionov}\ and\ \citenamefont
  {Mosel}(2003)}]{Larionov:2003av}%
  \BibitemOpen
  \bibfield  {author} {\bibinfo {author} {\bibfnamefont {A.~B.}\ \bibnamefont
  {Larionov}}\ and\ \bibinfo {author} {\bibfnamefont {U.}~\bibnamefont
  {Mosel}},\ }\href {\doibase 10.1016/j.nuclphysa.2003.08.005} {\bibfield
  {journal} {\bibinfo  {journal} {Nucl. Phys. A}\ }\textbf {\bibinfo {volume}
  {728}},\ \bibinfo {pages} {135} (\bibinfo {year} {2003})},\ \Eprint
  {http://arxiv.org/abs/nucl-th/0307035} {arXiv:nucl-th/0307035} \BibitemShut
  {NoStop}%
\bibitem [{\citenamefont {Uma~Maheswari}\ \emph {et~al.}(1998)\citenamefont
  {Uma~Maheswari}, \citenamefont {Fuchs}, \citenamefont {Faessler},
  \citenamefont {Sehn}, \citenamefont {Kosov},\ and\ \citenamefont
  {Wang}}]{UmaMaheswari:1997ig}%
  \BibitemOpen
  \bibfield  {author} {\bibinfo {author} {\bibfnamefont {V.~S.}\ \bibnamefont
  {Uma~Maheswari}}, \bibinfo {author} {\bibfnamefont {C.}~\bibnamefont
  {Fuchs}}, \bibinfo {author} {\bibfnamefont {A.}~\bibnamefont {Faessler}},
  \bibinfo {author} {\bibfnamefont {L.}~\bibnamefont {Sehn}}, \bibinfo {author}
  {\bibfnamefont {D.~S.}\ \bibnamefont {Kosov}}, \ and\ \bibinfo {author}
  {\bibfnamefont {Z.}~\bibnamefont {Wang}},\ }\href {\doibase
  10.1016/S0375-9474(97)00646-5} {\bibfield  {journal} {\bibinfo  {journal}
  {Nucl. Phys. A}\ }\textbf {\bibinfo {volume} {628}},\ \bibinfo {pages} {669}
  (\bibinfo {year} {1998})},\ \Eprint {http://arxiv.org/abs/nucl-th/9706004}
  {arXiv:nucl-th/9706004} \BibitemShut {NoStop}%
\bibitem [{\citenamefont {Liu}\ \emph {et~al.}(2002)\citenamefont {Liu},
  \citenamefont {Greco}, \citenamefont {Baran}, \citenamefont {Colonna},\ and\
  \citenamefont {Di~Toro}}]{Liu:2001iz}%
  \BibitemOpen
  \bibfield  {author} {\bibinfo {author} {\bibfnamefont {B.}~\bibnamefont
  {Liu}}, \bibinfo {author} {\bibfnamefont {V.}~\bibnamefont {Greco}}, \bibinfo
  {author} {\bibfnamefont {V.}~\bibnamefont {Baran}}, \bibinfo {author}
  {\bibfnamefont {M.}~\bibnamefont {Colonna}}, \ and\ \bibinfo {author}
  {\bibfnamefont {M.}~\bibnamefont {Di~Toro}},\ }\href {\doibase
  10.1103/PhysRevC.65.045201} {\bibfield  {journal} {\bibinfo  {journal} {Phys.
  Rev. C}\ }\textbf {\bibinfo {volume} {65}},\ \bibinfo {pages} {045201}
  (\bibinfo {year} {2002})},\ \Eprint {http://arxiv.org/abs/nucl-th/0112034}
  {arXiv:nucl-th/0112034} \BibitemShut {NoStop}%
\bibitem [{\citenamefont {Godbey}\ \emph {et~al.}(2022)\citenamefont {Godbey},
  \citenamefont {Zhang}, \citenamefont {Holt},\ and\ \citenamefont
  {Ko}}]{Godbey:2021tbt}%
  \BibitemOpen
  \bibfield  {author} {\bibinfo {author} {\bibfnamefont {K.}~\bibnamefont
  {Godbey}}, \bibinfo {author} {\bibfnamefont {Z.}~\bibnamefont {Zhang}},
  \bibinfo {author} {\bibfnamefont {J.~W.}\ \bibnamefont {Holt}}, \ and\
  \bibinfo {author} {\bibfnamefont {C.~M.}\ \bibnamefont {Ko}},\ }\href
  {\doibase 10.1016/j.physletb.2022.137134} {\bibfield  {journal} {\bibinfo
  {journal} {Phys. Lett. B}\ }\textbf {\bibinfo {volume} {829}},\ \bibinfo
  {pages} {137134} (\bibinfo {year} {2022})},\ \Eprint
  {http://arxiv.org/abs/2107.13384} {arXiv:2107.13384 [nucl-th]} \BibitemShut
  {NoStop}%
\bibitem [{\citenamefont {Kim}\ \emph {et~al.}(2022)\citenamefont {Kim},
  \citenamefont {Kim}, \citenamefont {Jeon},\ and\ \citenamefont
  {Lee}}]{Kim:2022sbj}%
  \BibitemOpen
  \bibfield  {author} {\bibinfo {author} {\bibfnamefont {M.}~\bibnamefont
  {Kim}}, \bibinfo {author} {\bibfnamefont {Y.}~\bibnamefont {Kim}}, \bibinfo
  {author} {\bibfnamefont {S.}~\bibnamefont {Jeon}}, \ and\ \bibinfo {author}
  {\bibfnamefont {C.~H.}\ \bibnamefont {Lee}},\ }\href {\doibase
  10.3390/universe8110564} {\bibfield  {journal} {\bibinfo  {journal}
  {Universe}\ }\textbf {\bibinfo {volume} {8}},\ \bibinfo {pages} {564}
  (\bibinfo {year} {2022})}\BibitemShut {NoStop}%
\bibitem [{\citenamefont {Kummer}\ \emph {et~al.}(2024)\citenamefont {Kummer},
  \citenamefont {Gallmeister},\ and\ \citenamefont {von
  Smekal}}]{Kummer:2023hvl}%
  \BibitemOpen
  \bibfield  {author} {\bibinfo {author} {\bibfnamefont {C.}~\bibnamefont
  {Kummer}}, \bibinfo {author} {\bibfnamefont {K.}~\bibnamefont {Gallmeister}},
  \ and\ \bibinfo {author} {\bibfnamefont {L.}~\bibnamefont {von Smekal}},\
  }\href {\doibase 10.1103/PhysRevC.109.054901} {\bibfield  {journal} {\bibinfo
   {journal} {Phys. Rev. C}\ }\textbf {\bibinfo {volume} {109}},\ \bibinfo
  {pages} {054901} (\bibinfo {year} {2024})},\ \Eprint
  {http://arxiv.org/abs/2309.09042} {arXiv:2309.09042 [nucl-th]} \BibitemShut
  {NoStop}%
\bibitem [{\citenamefont {Li}\ \emph {et~al.}(2025)\citenamefont {Li},
  \citenamefont {Wang}, \citenamefont {Zhang}, \citenamefont {Wang},
  \citenamefont {Pu}, \citenamefont {Ma},\ and\ \citenamefont
  {Chen}}]{Li:2025uku}%
  \BibitemOpen
  \bibfield  {author} {\bibinfo {author} {\bibfnamefont {X.}~\bibnamefont
  {Li}}, \bibinfo {author} {\bibfnamefont {S.~P.}\ \bibnamefont {Wang}},
  \bibinfo {author} {\bibfnamefont {Z.}~\bibnamefont {Zhang}}, \bibinfo
  {author} {\bibfnamefont {R.}~\bibnamefont {Wang}}, \bibinfo {author}
  {\bibfnamefont {J.}~\bibnamefont {Pu}}, \bibinfo {author} {\bibfnamefont
  {C.~W.}\ \bibnamefont {Ma}}, \ and\ \bibinfo {author} {\bibfnamefont {L.~W.}\
  \bibnamefont {Chen}},\ }\href@noop {} {\  (\bibinfo {year} {2025})},\ \Eprint
  {http://arxiv.org/abs/2509.21099} {arXiv:2509.21099 [nucl-th]} \BibitemShut
  {NoStop}%
\bibitem [{\citenamefont {Guo}\ and\ \citenamefont {Chen}(2025)}]{Guo:2025xie}%
  \BibitemOpen
  \bibfield  {author} {\bibinfo {author} {\bibfnamefont {W.~M.}\ \bibnamefont
  {Guo}}\ and\ \bibinfo {author} {\bibfnamefont {C.~H.}\ \bibnamefont {Chen}},\
  }\href {\doibase 10.1103/PhysRevC.111.024612} {\bibfield  {journal} {\bibinfo
   {journal} {Phys. Rev. C}\ }\textbf {\bibinfo {volume} {111}},\ \bibinfo
  {pages} {024612} (\bibinfo {year} {2025})}\BibitemShut {NoStop}%
\bibitem [{\citenamefont {Han}\ \emph {et~al.}(2025)\citenamefont {Han},
  \citenamefont {Shang}, \citenamefont {Zuo}, \citenamefont {Yong},\ and\
  \citenamefont {Li}}]{Han:2025hqm}%
  \BibitemOpen
  \bibfield  {author} {\bibinfo {author} {\bibfnamefont {S.~C.}\ \bibnamefont
  {Han}}, \bibinfo {author} {\bibfnamefont {X.~L.}\ \bibnamefont {Shang}},
  \bibinfo {author} {\bibfnamefont {W.}~\bibnamefont {Zuo}}, \bibinfo {author}
  {\bibfnamefont {G.~C.}\ \bibnamefont {Yong}}, \ and\ \bibinfo {author}
  {\bibfnamefont {A.}~\bibnamefont {Li}},\ }\href@noop {} {\  (\bibinfo {year}
  {2025})},\ \Eprint {http://arxiv.org/abs/2507.23476} {arXiv:2507.23476
  [nucl-th]} \BibitemShut {NoStop}%
\bibitem [{\citenamefont {Adamczewski-Musch}\ and\ \citenamefont {\textit{et
  al.}}(2020)}]{HADES:2020ver}%
  \BibitemOpen
  \bibfield  {author} {\bibinfo {author} {\bibfnamefont {J.}~\bibnamefont
  {Adamczewski-Musch}}\ and\ \bibinfo {author} {\bibnamefont {\textit{et al.}}}
  (\bibinfo {collaboration} {HADES Collaboration}),\ }\href {\doibase
  10.1140/epja/s10050-020-00237-2} {\bibfield  {journal} {\bibinfo  {journal}
  {Eur. Phys. J. A}\ }\textbf {\bibinfo {volume} {56}},\ \bibinfo {pages} {259}
  (\bibinfo {year} {2020})},\ \Eprint {http://arxiv.org/abs/2005.08774}
  {arXiv:2005.08774 [nucl-ex]} \BibitemShut {NoStop}%
\bibitem [{\citenamefont {Gaitanos}\ \emph {et~al.}(1999)\citenamefont
  {Gaitanos}, \citenamefont {Fuchs},\ and\ \citenamefont
  {Wolter}}]{Gaitanos:1998bu}%
  \BibitemOpen
  \bibfield  {author} {\bibinfo {author} {\bibfnamefont {T.}~\bibnamefont
  {Gaitanos}}, \bibinfo {author} {\bibfnamefont {C.}~\bibnamefont {Fuchs}}, \
  and\ \bibinfo {author} {\bibfnamefont {H.~H.}\ \bibnamefont {Wolter}},\
  }\href {\doibase 10.1016/S0375-9474(99)00002-0} {\bibfield  {journal}
  {\bibinfo  {journal} {Nucl. Phys. A}\ }\textbf {\bibinfo {volume} {650}},\
  \bibinfo {pages} {97} (\bibinfo {year} {1999})},\ \Eprint
  {http://arxiv.org/abs/nucl-th/9805025} {arXiv:nucl-th/9805025} \BibitemShut
  {NoStop}%
\bibitem [{\citenamefont {Zhang}\ and\ \citenamefont
  {Ko}(2018)}]{Zhang:2018ool}%
  \BibitemOpen
  \bibfield  {author} {\bibinfo {author} {\bibfnamefont {Z.}~\bibnamefont
  {Zhang}}\ and\ \bibinfo {author} {\bibfnamefont {C.~M.}\ \bibnamefont {Ko}},\
  }\href {\doibase 10.1103/PhysRevC.98.054614} {\bibfield  {journal} {\bibinfo
  {journal} {Phys. Rev. C}\ }\textbf {\bibinfo {volume} {98}},\ \bibinfo
  {pages} {054614} (\bibinfo {year} {2018})}\BibitemShut {NoStop}%
\bibitem [{\citenamefont {Walecka}(1974)}]{Walecka:1974qa}%
  \BibitemOpen
  \bibfield  {author} {\bibinfo {author} {\bibfnamefont {J.~D.}\ \bibnamefont
  {Walecka}},\ }\href {\doibase 10.1016/0003-4916(74)90208-5} {\bibfield
  {journal} {\bibinfo  {journal} {Annals Phys.}\ }\textbf {\bibinfo {volume}
  {83}},\ \bibinfo {pages} {491} (\bibinfo {year} {1974})}\BibitemShut
  {NoStop}%
\bibitem [{\citenamefont {Mao}\ \emph {et~al.}(1994{\natexlab{b}})\citenamefont
  {Mao}, \citenamefont {Li}, \citenamefont {Zhuo}, \citenamefont {Han},
  \citenamefont {Yu},\ and\ \citenamefont {Sano}}]{Mao:1994et}%
  \BibitemOpen
  \bibfield  {author} {\bibinfo {author} {\bibfnamefont {G.~J.}\ \bibnamefont
  {Mao}}, \bibinfo {author} {\bibfnamefont {Z.~X.}\ \bibnamefont {Li}},
  \bibinfo {author} {\bibfnamefont {Y.~Z.}\ \bibnamefont {Zhuo}}, \bibinfo
  {author} {\bibfnamefont {Y.~L.}\ \bibnamefont {Han}}, \bibinfo {author}
  {\bibfnamefont {Z.~Q.}\ \bibnamefont {Yu}}, \ and\ \bibinfo {author}
  {\bibfnamefont {M.}~\bibnamefont {Sano}},\ }\href {\doibase
  10.1007/BF01292373} {\bibfield  {journal} {\bibinfo  {journal} {Z. Phys. A}\
  }\textbf {\bibinfo {volume} {347}},\ \bibinfo {pages} {173} (\bibinfo {year}
  {1994}{\natexlab{b}})}\BibitemShut {NoStop}%
\bibitem [{\citenamefont {Li}\ \emph {et~al.}(2000)\citenamefont {Li},
  \citenamefont {Li},\ and\ \citenamefont {Mao}}]{Li:2000sha}%
  \BibitemOpen
  \bibfield  {author} {\bibinfo {author} {\bibfnamefont {Q.~F.}\ \bibnamefont
  {Li}}, \bibinfo {author} {\bibfnamefont {Z.~X.}\ \bibnamefont {Li}}, \ and\
  \bibinfo {author} {\bibfnamefont {G.~J.}\ \bibnamefont {Mao}},\ }\href
  {\doibase 10.1103/PhysRevC.62.014606} {\bibfield  {journal} {\bibinfo
  {journal} {Phys. Rev. C}\ }\textbf {\bibinfo {volume} {62}},\ \bibinfo
  {pages} {014606} (\bibinfo {year} {2000})},\ \Eprint
  {http://arxiv.org/abs/nucl-th/0005012} {arXiv:nucl-th/0005012} \BibitemShut
  {NoStop}%
\bibitem [{\citenamefont {Li}\ \emph {et~al.}(2004)\citenamefont {Li},
  \citenamefont {Li},\ and\ \citenamefont {Zhao}}]{Li:2003vd}%
  \BibitemOpen
  \bibfield  {author} {\bibinfo {author} {\bibfnamefont {Q.~F.}\ \bibnamefont
  {Li}}, \bibinfo {author} {\bibfnamefont {Z.~X.}\ \bibnamefont {Li}}, \ and\
  \bibinfo {author} {\bibfnamefont {E.~G.}\ \bibnamefont {Zhao}},\ }\href
  {\doibase 10.1103/PhysRevC.69.017601} {\bibfield  {journal} {\bibinfo
  {journal} {Phys. Rev. C}\ }\textbf {\bibinfo {volume} {69}},\ \bibinfo
  {pages} {017601} (\bibinfo {year} {2004})},\ \Eprint
  {http://arxiv.org/abs/nucl-th/0312098} {arXiv:nucl-th/0312098} \BibitemShut
  {NoStop}%
\bibitem [{\citenamefont {Li}\ and\ \citenamefont {Li}(2017)}]{Li:2016xix}%
  \BibitemOpen
  \bibfield  {author} {\bibinfo {author} {\bibfnamefont {Q.~F.}\ \bibnamefont
  {Li}}\ and\ \bibinfo {author} {\bibfnamefont {Z.~X.}\ \bibnamefont {Li}},\
  }\href {\doibase 10.1016/j.physletb.2017.09.013} {\bibfield  {journal}
  {\bibinfo  {journal} {Phys. Lett. B}\ }\textbf {\bibinfo {volume} {773}},\
  \bibinfo {pages} {557} (\bibinfo {year} {2017})},\ \Eprint
  {http://arxiv.org/abs/1610.00827} {arXiv:1610.00827 [nucl-th]} \BibitemShut
  {NoStop}%
\bibitem [{\citenamefont {Li}\ and\ \citenamefont {Li}(2019)}]{Li:2017pis}%
  \BibitemOpen
  \bibfield  {author} {\bibinfo {author} {\bibfnamefont {Q.~F.}\ \bibnamefont
  {Li}}\ and\ \bibinfo {author} {\bibfnamefont {Z.~X.}\ \bibnamefont {Li}},\
  }\href {\doibase 10.1007/s11433-018-9336-y} {\bibfield  {journal} {\bibinfo
  {journal} {Sci. China Phys. Mech. Astron.}\ }\textbf {\bibinfo {volume}
  {62}},\ \bibinfo {pages} {972011} (\bibinfo {year} {2019})},\ \Eprint
  {http://arxiv.org/abs/1712.02062} {arXiv:1712.02062 [nucl-th]} \BibitemShut
  {NoStop}%
\bibitem [{\citenamefont {Nan}\ \emph {et~al.}(2024)\citenamefont {Nan},
  \citenamefont {Li}, \citenamefont {Wang}, \citenamefont {Li},\ and\
  \citenamefont {Zuo}}]{Nan:2023gwp}%
  \BibitemOpen
  \bibfield  {author} {\bibinfo {author} {\bibfnamefont {M.~Z.}\ \bibnamefont
  {Nan}}, \bibinfo {author} {\bibfnamefont {P.~C.}\ \bibnamefont {Li}},
  \bibinfo {author} {\bibfnamefont {Y.~J.}\ \bibnamefont {Wang}}, \bibinfo
  {author} {\bibfnamefont {Q.~F.}\ \bibnamefont {Li}}, \ and\ \bibinfo {author}
  {\bibfnamefont {W.}~\bibnamefont {Zuo}},\ }\href {\doibase
  10.1140/epja/s10050-024-01349-9} {\bibfield  {journal} {\bibinfo  {journal}
  {Eur. Phys. J. A}\ }\textbf {\bibinfo {volume} {60}},\ \bibinfo {pages} {131}
  (\bibinfo {year} {2024})},\ \Eprint {http://arxiv.org/abs/2312.15716}
  {arXiv:2312.15716 [nucl-th]} \BibitemShut {NoStop}%
\bibitem [{\citenamefont {Nan}\ \emph {et~al.}(2025)\citenamefont {Nan},
  \citenamefont {Li}, \citenamefont {Zuo},\ and\ \citenamefont
  {Li}}]{Nan:2024ogc}%
  \BibitemOpen
  \bibfield  {author} {\bibinfo {author} {\bibfnamefont {M.~Z.}\ \bibnamefont
  {Nan}}, \bibinfo {author} {\bibfnamefont {P.~C.}\ \bibnamefont {Li}},
  \bibinfo {author} {\bibfnamefont {W.}~\bibnamefont {Zuo}}, \ and\ \bibinfo
  {author} {\bibfnamefont {Q.~F.}\ \bibnamefont {Li}},\ }\href {\doibase
  10.1088/1674-1137/add8fd} {\bibfield  {journal} {\bibinfo  {journal} {Chin.
  Phys. C}\ }\textbf {\bibinfo {volume} {49}},\ \bibinfo {pages} {094112}
  (\bibinfo {year} {2025})},\ \Eprint {http://arxiv.org/abs/2412.13497}
  {arXiv:2412.13497 [nucl-th]} \BibitemShut {NoStop}%
\bibitem [{\citenamefont {Gaitanos}\ \emph {et~al.}(2004)\citenamefont
  {Gaitanos}, \citenamefont {Di~Toro}, \citenamefont {Typel}, \citenamefont
  {Baran}, \citenamefont {Fuchs}, \citenamefont {Greco},\ and\ \citenamefont
  {Wolter}}]{Gaitanos:2003zg}%
  \BibitemOpen
  \bibfield  {author} {\bibinfo {author} {\bibfnamefont {T.}~\bibnamefont
  {Gaitanos}}, \bibinfo {author} {\bibfnamefont {M.}~\bibnamefont {Di~Toro}},
  \bibinfo {author} {\bibfnamefont {S.}~\bibnamefont {Typel}}, \bibinfo
  {author} {\bibfnamefont {V.}~\bibnamefont {Baran}}, \bibinfo {author}
  {\bibfnamefont {C.}~\bibnamefont {Fuchs}}, \bibinfo {author} {\bibfnamefont
  {V.}~\bibnamefont {Greco}}, \ and\ \bibinfo {author} {\bibfnamefont {H.~H.}\
  \bibnamefont {Wolter}},\ }\href {\doibase 10.1016/j.nuclphysa.2003.12.001}
  {\bibfield  {journal} {\bibinfo  {journal} {Nucl. Phys. A}\ }\textbf
  {\bibinfo {volume} {732}},\ \bibinfo {pages} {24} (\bibinfo {year} {2004})},\
  \Eprint {http://arxiv.org/abs/nucl-th/0309021} {arXiv:nucl-th/0309021}
  \BibitemShut {NoStop}%
\bibitem [{\citenamefont {Reichert}\ \emph {et~al.}(2023)\citenamefont
  {Reichert}, \citenamefont {Kittiratpattana}, \citenamefont {Li},
  \citenamefont {Steinheimer},\ and\ \citenamefont
  {Bleicher}}]{Reichert:2022qys}%
  \BibitemOpen
  \bibfield  {author} {\bibinfo {author} {\bibfnamefont {T.}~\bibnamefont
  {Reichert}}, \bibinfo {author} {\bibfnamefont {A.}~\bibnamefont
  {Kittiratpattana}}, \bibinfo {author} {\bibfnamefont {P.~C.}\ \bibnamefont
  {Li}}, \bibinfo {author} {\bibfnamefont {J.}~\bibnamefont {Steinheimer}}, \
  and\ \bibinfo {author} {\bibfnamefont {M.}~\bibnamefont {Bleicher}},\ }\href
  {\doibase 10.1088/1361-6471/acaffa} {\bibfield  {journal} {\bibinfo
  {journal} {J. Phys. G}\ }\textbf {\bibinfo {volume} {50}},\ \bibinfo {pages}
  {025104} (\bibinfo {year} {2023})},\ \Eprint
  {http://arxiv.org/abs/2208.10871} {arXiv:2208.10871 [nucl-th]} \BibitemShut
  {NoStop}%
\bibitem [{\citenamefont {Reichert}\ \emph {et~al.}(2021)\citenamefont
  {Reichert}, \citenamefont {Hillmann},\ and\ \citenamefont
  {Bleicher}}]{Reichert:2020uxs}%
  \BibitemOpen
  \bibfield  {author} {\bibinfo {author} {\bibfnamefont {T.}~\bibnamefont
  {Reichert}}, \bibinfo {author} {\bibfnamefont {P.}~\bibnamefont {Hillmann}},
  \ and\ \bibinfo {author} {\bibfnamefont {M.}~\bibnamefont {Bleicher}},\
  }\href {\doibase 10.1016/j.nuclphysa.2020.122058} {\bibfield  {journal}
  {\bibinfo  {journal} {Nucl. Phys. A}\ }\textbf {\bibinfo {volume} {1007}},\
  \bibinfo {pages} {122058} (\bibinfo {year} {2021})},\ \Eprint
  {http://arxiv.org/abs/2004.10539} {arXiv:2004.10539 [nucl-th]} \BibitemShut
  {NoStop}%
\bibitem [{\citenamefont {Brockmann}\ and\ \citenamefont {\textit{et
  al.}}(1984)}]{Brockmann:1984de}%
  \BibitemOpen
  \bibfield  {author} {\bibinfo {author} {\bibfnamefont {R.}~\bibnamefont
  {Brockmann}}\ and\ \bibinfo {author} {\bibnamefont {\textit{et al.}}},\
  }\href {\doibase 10.1103/PhysRevLett.53.2012} {\bibfield  {journal} {\bibinfo
   {journal} {Phys. Rev. Lett.}\ }\textbf {\bibinfo {volume} {53}},\ \bibinfo
  {pages} {2012} (\bibinfo {year} {1984})}\BibitemShut {NoStop}%
\bibitem [{\citenamefont {Wang}\ and\ \citenamefont {\textit{et
  al.}}(1996)}]{Wang:1996zz}%
  \BibitemOpen
  \bibfield  {author} {\bibinfo {author} {\bibfnamefont {S.}~\bibnamefont
  {Wang}}\ and\ \bibinfo {author} {\bibnamefont {\textit{et al.}}},\ }\href
  {\doibase 10.1103/PhysRevLett.76.3911} {\bibfield  {journal} {\bibinfo
  {journal} {Phys. Rev. Lett.}\ }\textbf {\bibinfo {volume} {76}},\ \bibinfo
  {pages} {3911} (\bibinfo {year} {1996})}\BibitemShut {NoStop}%
\bibitem [{\citenamefont {Andronic}\ \emph {et~al.}(2006)\citenamefont
  {Andronic}, \citenamefont {Braun~Munzinger},\ and\ \citenamefont
  {Stachel}}]{Andronic:2005yp}%
  \BibitemOpen
  \bibfield  {author} {\bibinfo {author} {\bibfnamefont {A.}~\bibnamefont
  {Andronic}}, \bibinfo {author} {\bibfnamefont {P.}~\bibnamefont
  {Braun~Munzinger}}, \ and\ \bibinfo {author} {\bibfnamefont {J.}~\bibnamefont
  {Stachel}},\ }\href {\doibase 10.1016/j.nuclphysa.2006.03.012} {\bibfield
  {journal} {\bibinfo  {journal} {Nucl. Phys. A}\ }\textbf {\bibinfo {volume}
  {772}},\ \bibinfo {pages} {167} (\bibinfo {year} {2006})},\ \Eprint
  {http://arxiv.org/abs/nucl-th/0511071} {arXiv:nucl-th/0511071} \BibitemShut
  {NoStop}%
\bibitem [{\citenamefont {Sen}(2021)}]{Sen:2020edi}%
  \BibitemOpen
  \bibfield  {author} {\bibinfo {author} {\bibfnamefont {D.}~\bibnamefont
  {Sen}},\ }\href {\doibase 10.1088/1361-6471/abcb9e} {\bibfield  {journal}
  {\bibinfo  {journal} {J. Phys. G}\ }\textbf {\bibinfo {volume} {48}},\
  \bibinfo {pages} {025201} (\bibinfo {year} {2021})},\ \Eprint
  {http://arxiv.org/abs/2011.09785} {arXiv:2011.09785 [nucl-th]} \BibitemShut
  {NoStop}%
\bibitem [{\citenamefont {Miyatsu}\ \emph {et~al.}(2022)\citenamefont
  {Miyatsu}, \citenamefont {Cheoun},\ and\ \citenamefont
  {Saito}}]{Miyatsu:2022wuy}%
  \BibitemOpen
  \bibfield  {author} {\bibinfo {author} {\bibfnamefont {T.}~\bibnamefont
  {Miyatsu}}, \bibinfo {author} {\bibfnamefont {M.~K.}\ \bibnamefont {Cheoun}},
  \ and\ \bibinfo {author} {\bibfnamefont {K.}~\bibnamefont {Saito}},\ }\href
  {\doibase 10.3847/1538-4357/ac5f40} {\bibfield  {journal} {\bibinfo
  {journal} {Astrophys. J.}\ }\textbf {\bibinfo {volume} {929}},\ \bibinfo
  {pages} {82} (\bibinfo {year} {2022})},\ \Eprint
  {http://arxiv.org/abs/2202.06468} {arXiv:2202.06468 [nucl-th]} \BibitemShut
  {NoStop}%
\bibitem [{\citenamefont {Sun}\ \emph {et~al.}(2023)\citenamefont {Sun},
  \citenamefont {Miao}, \citenamefont {Sun},\ and\ \citenamefont
  {Li}}]{Sun:2022yor}%
  \BibitemOpen
  \bibfield  {author} {\bibinfo {author} {\bibfnamefont {X.~D.}\ \bibnamefont
  {Sun}}, \bibinfo {author} {\bibfnamefont {Z.~Q.}\ \bibnamefont {Miao}},
  \bibinfo {author} {\bibfnamefont {B.~Y.}\ \bibnamefont {Sun}}, \ and\
  \bibinfo {author} {\bibfnamefont {A.}~\bibnamefont {Li}},\ }\href {\doibase
  10.3847/1538-4357/ac9d9a} {\bibfield  {journal} {\bibinfo  {journal}
  {Astrophys. J.}\ }\textbf {\bibinfo {volume} {942}},\ \bibinfo {pages} {55}
  (\bibinfo {year} {2023})},\ \Eprint {http://arxiv.org/abs/2205.10631}
  {arXiv:2205.10631 [astro-ph.HE]} \BibitemShut {NoStop}%
\bibitem [{\citenamefont {Roca~Maza}\ \emph {et~al.}(2011)\citenamefont
  {Roca~Maza}, \citenamefont {Vinas}, \citenamefont {Centelles}, \citenamefont
  {Ring},\ and\ \citenamefont {Schuck}}]{Roca-Maza:2011alv}%
  \BibitemOpen
  \bibfield  {author} {\bibinfo {author} {\bibfnamefont {X.}~\bibnamefont
  {Roca~Maza}}, \bibinfo {author} {\bibfnamefont {X.}~\bibnamefont {Vinas}},
  \bibinfo {author} {\bibfnamefont {M.}~\bibnamefont {Centelles}}, \bibinfo
  {author} {\bibfnamefont {P.}~\bibnamefont {Ring}}, \ and\ \bibinfo {author}
  {\bibfnamefont {P.}~\bibnamefont {Schuck}},\ }\href {\doibase
  10.1103/PhysRevC.84.054309} {\bibfield  {journal} {\bibinfo  {journal} {Phys.
  Rev. C}\ }\textbf {\bibinfo {volume} {84}},\ \bibinfo {pages} {054309}
  (\bibinfo {year} {2011})},\ \bibinfo {note} {[Erratum: Phys.Rev.C 93, 069905
  (2016)]},\ \Eprint {http://arxiv.org/abs/1110.2311} {arXiv:1110.2311
  [nucl-th]} \BibitemShut {NoStop}%
\bibitem [{\citenamefont {Machleidt}(1989)}]{Machleidt:1989tm}%
  \BibitemOpen
  \bibfield  {author} {\bibinfo {author} {\bibfnamefont {R.}~\bibnamefont
  {Machleidt}},\ }\href@noop {} {\bibfield  {journal} {\bibinfo  {journal}
  {Adv. Nucl. Phys.}\ }\textbf {\bibinfo {volume} {19}},\ \bibinfo {pages}
  {189} (\bibinfo {year} {1989})}\BibitemShut {NoStop}%
\bibitem [{\citenamefont {Zabari}\ \emph
  {et~al.}(2019{\natexlab{a}})\citenamefont {Zabari}, \citenamefont {Kubis},\
  and\ \citenamefont {W\'ojcik}}]{Zabari:2018tjk}%
  \BibitemOpen
  \bibfield  {author} {\bibinfo {author} {\bibfnamefont {N.}~\bibnamefont
  {Zabari}}, \bibinfo {author} {\bibfnamefont {S.}~\bibnamefont {Kubis}}, \
  and\ \bibinfo {author} {\bibfnamefont {W.}~\bibnamefont {W\'ojcik}},\ }\href
  {\doibase 10.1103/PhysRevC.99.035209} {\bibfield  {journal} {\bibinfo
  {journal} {Phys. Rev. C}\ }\textbf {\bibinfo {volume} {99}},\ \bibinfo
  {pages} {035209} (\bibinfo {year} {2019}{\natexlab{a}})},\ \Eprint
  {http://arxiv.org/abs/1809.03420} {arXiv:1809.03420 [nucl-th]} \BibitemShut
  {NoStop}%
\bibitem [{\citenamefont {Zabari}\ \emph
  {et~al.}(2019{\natexlab{b}})\citenamefont {Zabari}, \citenamefont {Kubis},\
  and\ \citenamefont {W\'ojcik}}]{Zabari:2019clk}%
  \BibitemOpen
  \bibfield  {author} {\bibinfo {author} {\bibfnamefont {N.}~\bibnamefont
  {Zabari}}, \bibinfo {author} {\bibfnamefont {S.}~\bibnamefont {Kubis}}, \
  and\ \bibinfo {author} {\bibfnamefont {W.}~\bibnamefont {W\'ojcik}},\ }\href
  {\doibase 10.1103/PhysRevC.100.015808} {\bibfield  {journal} {\bibinfo
  {journal} {Phys. Rev. C}\ }\textbf {\bibinfo {volume} {100}},\ \bibinfo
  {pages} {015808} (\bibinfo {year} {2019}{\natexlab{b}})},\ \Eprint
  {http://arxiv.org/abs/1908.00476} {arXiv:1908.00476 [nucl-th]} \BibitemShut
  {NoStop}%
\bibitem [{\citenamefont {Yang}\ \emph {et~al.}(2025)\citenamefont {Yang},
  \citenamefont {Sun}, \citenamefont {Cui}, \citenamefont {Liu}, \citenamefont
  {Li}, \citenamefont {Zhao},\ and\ \citenamefont {Zhang}}]{Yang:2025aia}%
  \BibitemOpen
  \bibfield  {author} {\bibinfo {author} {\bibfnamefont {J.}~\bibnamefont
  {Yang}}, \bibinfo {author} {\bibfnamefont {M.}~\bibnamefont {Sun}}, \bibinfo
  {author} {\bibfnamefont {Y.}~\bibnamefont {Cui}}, \bibinfo {author}
  {\bibfnamefont {Y.}~\bibnamefont {Liu}}, \bibinfo {author} {\bibfnamefont
  {Z.}~\bibnamefont {Li}}, \bibinfo {author} {\bibfnamefont {K.}~\bibnamefont
  {Zhao}}, \ and\ \bibinfo {author} {\bibfnamefont {Y.}~\bibnamefont {Zhang}},\
  }\href@noop {} {\  (\bibinfo {year} {2025})},\ \Eprint
  {http://arxiv.org/abs/2506.17973} {arXiv:2506.17973 [nucl-th]} \BibitemShut
  {NoStop}%
\bibitem [{\citenamefont {Li}(2004)}]{Li:2004zi}%
  \BibitemOpen
  \bibfield  {author} {\bibinfo {author} {\bibfnamefont {B.~A.}\ \bibnamefont
  {Li}},\ }\href {\doibase 10.1103/PhysRevC.69.064602} {\bibfield  {journal}
  {\bibinfo  {journal} {Phys. Rev. C}\ }\textbf {\bibinfo {volume} {69}},\
  \bibinfo {pages} {064602} (\bibinfo {year} {2004})},\ \Eprint
  {http://arxiv.org/abs/nucl-th/0404040} {arXiv:nucl-th/0404040} \BibitemShut
  {NoStop}%
\bibitem [{\citenamefont {Zhang}\ \emph {et~al.}(2014)\citenamefont {Zhang},
  \citenamefont {Tsang}, \citenamefont {Li},\ and\ \citenamefont
  {Liu}}]{Zhang:2014sva}%
  \BibitemOpen
  \bibfield  {author} {\bibinfo {author} {\bibfnamefont {Y.~X.}\ \bibnamefont
  {Zhang}}, \bibinfo {author} {\bibfnamefont {M.~B.}\ \bibnamefont {Tsang}},
  \bibinfo {author} {\bibfnamefont {Z.~X.}\ \bibnamefont {Li}}, \ and\ \bibinfo
  {author} {\bibfnamefont {H.}~\bibnamefont {Liu}},\ }\href {\doibase
  10.1016/j.physletb.2014.03.030} {\bibfield  {journal} {\bibinfo  {journal}
  {Phys. Lett. B}\ }\textbf {\bibinfo {volume} {732}},\ \bibinfo {pages} {186}
  (\bibinfo {year} {2014})},\ \Eprint {http://arxiv.org/abs/1402.3790}
  {arXiv:1402.3790 [nucl-th]} \BibitemShut {NoStop}%
\bibitem [{\citenamefont {Coupland}\ \emph {et~al.}(2016)\citenamefont
  {Coupland} \emph {et~al.}}]{Coupland:2014gya}%
  \BibitemOpen
  \bibfield  {author} {\bibinfo {author} {\bibfnamefont {D.~D.~S.}\
  \bibnamefont {Coupland}} \emph {et~al.},\ }\href {\doibase
  10.1103/PhysRevC.94.011601} {\bibfield  {journal} {\bibinfo  {journal} {Phys.
  Rev. C}\ }\textbf {\bibinfo {volume} {94}},\ \bibinfo {pages} {011601}
  (\bibinfo {year} {2016})},\ \Eprint {http://arxiv.org/abs/1406.4546}
  {arXiv:1406.4546 [nucl-ex]} \BibitemShut {NoStop}%
\bibitem [{\citenamefont {Su}\ \emph {et~al.}(2017)\citenamefont {Su},
  \citenamefont {Zhu}, \citenamefont {Huang}, \citenamefont {Xie},\ and\
  \citenamefont {Zhang}}]{Su:2017vwn}%
  \BibitemOpen
  \bibfield  {author} {\bibinfo {author} {\bibfnamefont {J.}~\bibnamefont
  {Su}}, \bibinfo {author} {\bibfnamefont {L.}~\bibnamefont {Zhu}}, \bibinfo
  {author} {\bibfnamefont {C.~Y.}\ \bibnamefont {Huang}}, \bibinfo {author}
  {\bibfnamefont {W.~J.}\ \bibnamefont {Xie}}, \ and\ \bibinfo {author}
  {\bibfnamefont {F.~S.}\ \bibnamefont {Zhang}},\ }\href {\doibase
  10.1103/PhysRevC.96.024601} {\bibfield  {journal} {\bibinfo  {journal} {Phys.
  Rev. C}\ }\textbf {\bibinfo {volume} {96}},\ \bibinfo {pages} {024601}
  (\bibinfo {year} {2017})}\BibitemShut {NoStop}%
\bibitem [{\citenamefont {Zhuang}\ \emph {et~al.}(1994)\citenamefont {Zhuang},
  \citenamefont {Hufner},\ and\ \citenamefont {Klevansky}}]{Zhuang:1994dw}%
  \BibitemOpen
  \bibfield  {author} {\bibinfo {author} {\bibfnamefont {P.}~\bibnamefont
  {Zhuang}}, \bibinfo {author} {\bibfnamefont {J.}~\bibnamefont {Hufner}}, \
  and\ \bibinfo {author} {\bibfnamefont {S.~P.}\ \bibnamefont {Klevansky}},\
  }\href {\doibase 10.1016/0375-9474(94)90743-9} {\bibfield  {journal}
  {\bibinfo  {journal} {Nucl. Phys. A}\ }\textbf {\bibinfo {volume} {576}},\
  \bibinfo {pages} {525} (\bibinfo {year} {1994})}\BibitemShut {NoStop}%
\bibitem [{\citenamefont {Chaudhuri}\ \emph {et~al.}(2019)\citenamefont
  {Chaudhuri}, \citenamefont {Ghosh}, \citenamefont {Sarkar},\ and\
  \citenamefont {Roy}}]{Chaudhuri:2019lbw}%
  \BibitemOpen
  \bibfield  {author} {\bibinfo {author} {\bibfnamefont {N.}~\bibnamefont
  {Chaudhuri}}, \bibinfo {author} {\bibfnamefont {S.}~\bibnamefont {Ghosh}},
  \bibinfo {author} {\bibfnamefont {S.}~\bibnamefont {Sarkar}}, \ and\ \bibinfo
  {author} {\bibfnamefont {P.}~\bibnamefont {Roy}},\ }\href {\doibase
  10.1103/PhysRevD.99.116025} {\bibfield  {journal} {\bibinfo  {journal} {Phys.
  Rev. D}\ }\textbf {\bibinfo {volume} {99}},\ \bibinfo {pages} {116025}
  (\bibinfo {year} {2019})},\ \Eprint {http://arxiv.org/abs/1907.03990}
  {arXiv:1907.03990 [nucl-th]} \BibitemShut {NoStop}%
\bibitem [{\citenamefont {Theis}\ \emph {et~al.}(1983)\citenamefont {Theis},
  \citenamefont {Graebner}, \citenamefont {Buchwald}, \citenamefont {Maruhn},
  \citenamefont {Greiner}, \citenamefont {Stoecker},\ and\ \citenamefont
  {Polonyi}}]{Theis:1983egm}%
  \BibitemOpen
  \bibfield  {author} {\bibinfo {author} {\bibfnamefont {J.}~\bibnamefont
  {Theis}}, \bibinfo {author} {\bibfnamefont {G.}~\bibnamefont {Graebner}},
  \bibinfo {author} {\bibfnamefont {G.}~\bibnamefont {Buchwald}}, \bibinfo
  {author} {\bibfnamefont {J.~A.}\ \bibnamefont {Maruhn}}, \bibinfo {author}
  {\bibfnamefont {W.}~\bibnamefont {Greiner}}, \bibinfo {author} {\bibfnamefont
  {H.}~\bibnamefont {Stoecker}}, \ and\ \bibinfo {author} {\bibfnamefont
  {J.}~\bibnamefont {Polonyi}},\ }\href {\doibase 10.1103/PhysRevD.28.2286}
  {\bibfield  {journal} {\bibinfo  {journal} {Phys. Rev. D}\ }\textbf {\bibinfo
  {volume} {28}},\ \bibinfo {pages} {2286} (\bibinfo {year}
  {1983})}\BibitemShut {NoStop}%
\bibitem [{\citenamefont {Freedman}(1977)}]{Freedman:1977fd}%
  \BibitemOpen
  \bibfield  {author} {\bibinfo {author} {\bibfnamefont {R.~A.}\ \bibnamefont
  {Freedman}},\ }\href {\doibase 10.1016/0370-2693(77)90242-8} {\bibfield
  {journal} {\bibinfo  {journal} {Phys. Lett. B}\ }\textbf {\bibinfo {volume}
  {71}},\ \bibinfo {pages} {369} (\bibinfo {year} {1977})}\BibitemShut
  {NoStop}%
\bibitem [{\citenamefont {Glendenning}(1987)}]{Glendenning:1986ib}%
  \BibitemOpen
  \bibfield  {author} {\bibinfo {author} {\bibfnamefont {N.~K.}\ \bibnamefont
  {Glendenning}},\ }\href {\doibase 10.1016/0370-2693(87)90999-3} {\bibfield
  {journal} {\bibinfo  {journal} {Phys. Lett. B}\ }\textbf {\bibinfo {volume}
  {185}},\ \bibinfo {pages} {275} (\bibinfo {year} {1987})}\BibitemShut
  {NoStop}%
\bibitem [{\citenamefont {Bellwied}\ \emph {et~al.}(2015)\citenamefont
  {Bellwied}, \citenamefont {Borsanyi}, \citenamefont {Fodor}, \citenamefont
  {G{\"u}nther}, \citenamefont {Katz}, \citenamefont {Ratti},\ and\
  \citenamefont {Szabo}}]{Bellwied:2015rza}%
  \BibitemOpen
  \bibfield  {author} {\bibinfo {author} {\bibfnamefont {R.}~\bibnamefont
  {Bellwied}}, \bibinfo {author} {\bibfnamefont {S.}~\bibnamefont {Borsanyi}},
  \bibinfo {author} {\bibfnamefont {Z.}~\bibnamefont {Fodor}}, \bibinfo
  {author} {\bibfnamefont {J.}~\bibnamefont {G{\"u}nther}}, \bibinfo {author}
  {\bibfnamefont {S.~D.}\ \bibnamefont {Katz}}, \bibinfo {author}
  {\bibfnamefont {C.}~\bibnamefont {Ratti}}, \ and\ \bibinfo {author}
  {\bibfnamefont {K.~K.}\ \bibnamefont {Szabo}},\ }\href {\doibase
  10.1016/j.physletb.2015.11.011} {\bibfield  {journal} {\bibinfo  {journal}
  {Phys. Lett. B}\ }\textbf {\bibinfo {volume} {751}},\ \bibinfo {pages} {559}
  (\bibinfo {year} {2015})},\ \Eprint {http://arxiv.org/abs/1507.07510}
  {arXiv:1507.07510 [hep-lat]} \BibitemShut {NoStop}%
\bibitem [{\citenamefont {Li}\ \emph {et~al.}(1993)\citenamefont {Li},
  \citenamefont {Zhuo}, \citenamefont {Gu}, \citenamefont {Sun}, \citenamefont
  {Yu},\ and\ \citenamefont {Sano}}]{Li:1993muy}%
  \BibitemOpen
  \bibfield  {author} {\bibinfo {author} {\bibfnamefont {Z.}~\bibnamefont
  {Li}}, \bibinfo {author} {\bibfnamefont {Y.}~\bibnamefont {Zhuo}}, \bibinfo
  {author} {\bibfnamefont {Y.}~\bibnamefont {Gu}}, \bibinfo {author}
  {\bibfnamefont {Z.}~\bibnamefont {Sun}}, \bibinfo {author} {\bibfnamefont
  {Z.}~\bibnamefont {Yu}}, \ and\ \bibinfo {author} {\bibfnamefont
  {M.}~\bibnamefont {Sano}},\ }\href {\doibase 10.1016/0375-9474(93)90262-V}
  {\bibfield  {journal} {\bibinfo  {journal} {Nucl. Phys. A}\ }\textbf
  {\bibinfo {volume} {559}},\ \bibinfo {pages} {603} (\bibinfo {year}
  {1993})}\BibitemShut {NoStop}%
\end{thebibliography}%
	
\end{document}